\documentclass[sigconf]{acmart}

\usepackage{booktabs} % For formal tables
\usepackage{subfig}
\usepackage{hhline}
\usepackage{array}
\usepackage{afterpage}
\usepackage{placeins}
\usepackage{color}

\def\changed#1{{#1}}
\AtBeginDocument{%
  \providecommand\BibTeX{{%
    \normalfont B\kern-0.5em{\scshape i\kern-0.25em b}\kern-0.8em\TeX}}}

\copyrightyear{2022} 
\acmYear{2022} 
\setcopyright{rightsretained} 
\acmConference[CHI '22]{CHI Conference on Human Factors in Computing
Systems}{April 29-May 5, 2022}{New Orleans, LA, USA}
\acmBooktitle{CHI Conference on Human Factors in Computing Systems
(CHI '22), April 29-May 5, 2022, New Orleans, LA, USA}
\acmDOI{10.1145/3491102.3517605}
\acmISBN{978-1-4503-9157-3/22/04}

\begin{document}

\title[Understanding AR Activism]{Understanding AR Activism: 
An Interview Study with Creators of Augmented Reality Experiences for Social Change}

%\author{Anonymous}
% \affiliation{%

\author{Rafael M.L. Silva}
\affiliation{
\institution{University of Washington}
\city{Seattle}
\state{WA}
\country{USA}
}
\email{rafaelsi@uw.edu}

\author{Erica Principe Cruz}
\affiliation{
\institution{Carnegie Mellon University}
\city{Pittsburgh}
\state{PA}
\country{United States}
}
\email{ecruz@cs.cmu.edu}

\author{Daniela K. Rosner}
\affiliation{
\institution{University of Washington}
\city{Seattle}
\state{WA}
\country{USA}
}
\email{dkrosner@uw.edu}

\author{Dayton Kelly}
\affiliation{
\institution{University of Washington}
\city{Seattle}
\state{WA}
\country{USA}
}
\email{bydkelly@gmail.com}

\author{Andrés Monroy-Hernández}
%\authornote{Also with Princeton University, Princeton, NJ, USA.}
\affiliation{
\institution{Snap Inc. \& Princeton University}
\city{Princeton}
\state{NJ}
\country{USA}
%\city{Seattle}
%\state{WA}
%\country{USA}
}
\email{andresmh@princeton.edu}

%\affiliation{
%\institution{Computer Science, Princeton University}
%\city{Princeton}
%\state{NJ}
%\country{USA}
%}

\author{Fannie Liu}
\affiliation{
\institution{Snap Inc.}
\city{New York}
\state{NY}
\country{USA}
}
\email{fannie@snap.com}

\renewcommand{\shortauthors}{Silva, et al.}

\begin{abstract}
The rise of consumer augmented reality (AR) technology has opened up new possibilities for interventions intended to disrupt and subvert cultural conventions. From defacing corporate logos to erecting geofenced digital monuments, more and more people are creating AR experiences for social causes. We sought to understand this new form of activism, including why people use AR for these purposes, opportunities and challenges in using it, and how well it can support activist goals. We conducted semi-structured interviews with twenty people involved in projects that used AR for a social cause across six different countries. We found that AR can overcome physical world limitations of activism to convey immersive, multilayered narratives that aim to reveal invisible histories and perspectives. At the same time, people experienced challenges in creating, maintaining, and distributing their AR experiences to audiences. We discuss open questions and opportunities for creating AR tools and experiences for social change.

\end{abstract}




\maketitle

\section{Introduction}

Augmented Reality (AR) has followed an adoption pattern that resembles prior technologies, i.e., moving from the lab to the home, and onto the public realm. Most of today's information technologies were first embraced by scientists, then consumers and activists. For example, the internet and the web were initially conceived and adopted by scientists at DARPA and CERN, respectively. Decades later, ordinary people embraced these technologies for work, school, and social life. Upon reaching mass adoption, activists used them to spread political opinions and help mobilize protests, among other activities. Scholars describing this pattern of technological adoption have argued that everyday uses of technology are prerequisites to its use for activism. For example, Zuckerman \cite{zuckerman_2007} posited that mundane online activities, such as posting photos of cute cats on the web, help people gain fluency with the medium and subsequently enable them to use it for activism. Furthermore, they argued that these pervasive uses make censorship a politically harder calculation for authoritarian regimes. 

Although still in early stages, AR has, too, grown into a platform for activism. Nearly half a century after Sutherland's 1968 AR prototypes \cite{sutherland1968head}, AR has become widely adopted through consumer applications like Snapchat's ``filters'' and location-based games like Pokémon Go.
More recently, AR creators have begun using the technology for activism.
For example, the ``Whole Story'' project (Figure 1A) created a GPS-enabled app that activates AR monuments of influential women at historical locations to address gender gap representation~\cite{WholeStoryProject}. Similarly, the ``Hack de Patria'' project (Figure 1B) exposed attempts to defraud voting ballots and overlaid subversive messages over the political propaganda of the Maduro regime during the 2015 Venezuelan elections~\cite{HackDaPatria}.

\begin{figure*}
    \centering
    \includegraphics[width=0.9\textwidth]{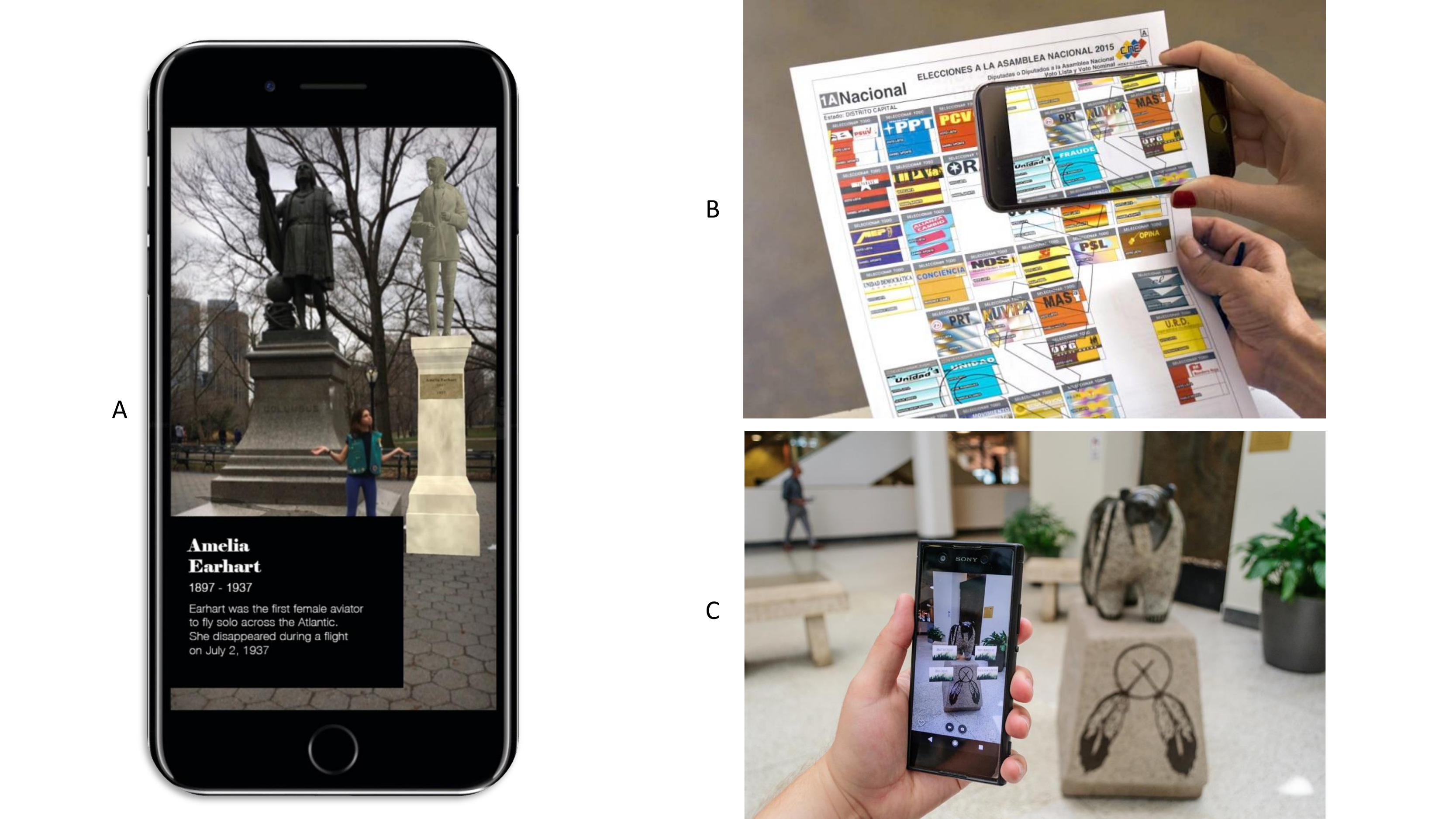}
    \caption{Examples of AR Activism projects: (A) Viewfinder of ``Whole Story'' app featuring a person standing between physical and virtual AR statues \changed{that highlight gaps in gender representation}, (B) ``Hack de Patria'' project augmenting voting ballots to expose attempts to mislead voters, and (C) ``SweetgrassAR'' augmenting a symbolic sculpture with stories about Indigenous knowledge.}
    \Description{Three examples of AR Activism projects augmenting places, voting ballots, and sculptures.}
\end{figure*}

AR is particularly interesting for ``activist tech'' research because the technology's core features are uniquely aligned with key activist practices. For instance, AR creators and activists are motivated by a desire to change the world and help people see it in new ways. Activists aim to alter the physical world and people's perceptions of it according to a social cause. Similarly, AR creators change the way people perceive the world by augmenting what people see and hear. Furthermore, activists are often tightly connected with the places and physical things that relate to their causes. AR experiences are likewise tightly coupled to the physical world because they rely on the environment's physical attributes to detect and augment objects, surfaces, and even plants, animals, and humans. Additionally, unlike other digital technologies, like social media or virtual reality, AR moves away from the illusion of digital dualism \cite{jurgenson_2011}, i.e., the idea that the online world is somehow separate from the ``real world.'' AR breaks that dualism and deeply builds on the interconnectedness of the physical and digital world. In AR, Barlow's declaration of independence for ``cyberspace'' becomes a fight for independence in the physical world itself, often a core goal of activists \cite{barlow_2018}.

At the same time, we know little about AR activism, including what influences people's decisions to use the technology for social change, opportunities and challenges in their subsequent experiences, and their ability to achieve their activist goals. To answer these questions, we conducted 60-minute interviews with twenty  people who participated in AR activism projects (with project start dates ranging from 2009-2021), based in six different countries. Our qualitative analysis of these interviews elicited the following key insights: (1) participants' \textbf{decisions} to use AR is driven by the possibility \textbf{to create rich multilayered narratives}, where they perceived \textbf{lower technical barriers and physical risks} to convey those narratives in the physical world. At the same time, their decisions were sometimes stymied by uncertainties around short- and long-term legal and ethical consequences.
Furthermore, (2) participants identified \textbf{immersion} as a central means of changing people's perceptions and reshaping the world due to its potential for embodied storytelling, and \textbf{interdisciplinary collaboration} as key to achieving high quality immersive experiences. On the other hand, the AR platforms and collaborations needed to create those experiences can come with their own challenges around resource management and expertise.
Lastly, (3) we found that participants' \textbf{perceptions of AR's impact} on their projects involved eliciting a strong \textbf{emotional and physical reaction} among their audiences. However, they struggled to get a broader reach due to challenges with the distribution and adoption of the technology.

We conclude with a set of open questions and opportunities around supporting multidisciplinary teams and communities for AR activism content creation, expanding activist-control by attending to maintenance and proprietary decision-making, addressing concerns around broader issues of discoverability and accountability in AR, and realizing radical transformations that support minoritarian groups.

\section{Related Work}
To set the stage for our study of AR activism and its creators, we turn to the variety of work on social and civic change that people enact with and around immersive technology. This scholarship and practice involves two central bodies of work: (1) digital design practices for activism, (2) AR-specific social and organizational change. We review these areas in the sections that follow and outline core questions they raise for our analysis.

\subsection{Design and Digital Activism}
Understanding activism and immersive technology involves comprehending the specific speculative practices and planning through which people engage with notions of social or political transformation, a topic commonly termed design activism \cite{fox2020accounting}. Design activism encompasses a diverse body of work at the intersection of urban studies, social movement theory, and critical design studies \cite{wylie2020promise, markussen2013disruptive, le2016designing}. To date, this scholarship has examined a range of phenomena around the conditions for organizing social and political change, including the influence of ``counternarrative'' \cite{fuad2013design}, the design-specific contours of community organizing \cite{thorpe2012architecture,tran2019gets,von2022making}, and the importance of communal accountability and care \cite{criado2020anthropology,fox2020accounting}. One crucial facet of this work concerns the specific forms of intervention that rely on digital platforms and services. Reviewing this literature, George and Leidner define \textit{digital activism} as ``digitally mediated social activism,'' where social activism includes ``taking action to create social change.'' According to this growing body of scholarship, people engage in a range of digital activist actions at different levels of engagement. For example, ``clicktivism'' at the lowest level includes ``liking'' a social media post, while ``hacktivism'' at the highest ``gladiatorial'' level involves hacking for a social/political goal~\cite{george2019clicktivism}.

Social media, such as Twitter and Facebook, are perhaps the most explored digital media for activism. While George and Leidner suggest that social media can be used at all levels of engagement~\cite{george2019clicktivism}, a number of HCI researchers have highlighted their ability to coordinate and broadcast messages. By quickly disseminating information to a large number of people, social media can help raise awareness, spark discussion, and organize both online and offline movements around violence~\cite{monroy2013new}, corruption~\cite{tufekci2012social}, accessibility~\cite{li2018slacktivists}, racial inequality~\cite{de2016social}, economic inequality~\cite{gleason2013occupy}, sexual harassment~\cite{dimond2013hollaback}, and more. For marginalized groups in particular, it can play an important role in democratizing their voices and increasing feelings of empowerment around their identities~\cite{dimond2013hollaback,liu2017selfies,li2018slacktivists}.

At the same time, several scholars have questioned the impact of digital activism, especially at the lowest level of ``clicktivism.'' Such online actions have been criticized as ``slacktivism'' for being low-cost, low-risk, and done in order to feel good despite little to no real-world effects~\cite{morozov:bravenewworld}. Others describe its main purpose as lowering the barrier for coordination, with primary activist actions happening in the physical world, such as protests~\cite{bennett2012logic,hirsch2005txtmob}. While recent work suggests that even low-level actions can lead to or correlate with real-world impact~\cite{lee2013does,milovsevic2017civic}, physically ``putting your body on the line'' is still considered highly meaningful among activists~\cite{li2018slacktivists}.

\textit{``Hybrid activism''} is an emerging form of activism that blends online and offline action~\cite{milovsevic2017civic,li2018slacktivists}. This activity can include using online tools to support offline action (e.g., coordinating a protest on social media) as well as synchronous online and offline action (e.g., attending a protest through telepresence~\cite{li2018slacktivists}). Recent activist actions suggest the emergence of another style of hybrid activism: using digital media \textit{in} the physical world for social change, rather than the two existing separately. For example, recent HCI research has investigated Internet of Things devices, which equip physical spaces with sensor technologies, and location-based mobile technologies for menstrual resource accessibility~\cite{fox2018beyond}, civic engagement~\cite{salim2015urban,paulos2009citizen}, public expression~\cite{kuznetsov2010wallbots}, and sustainability~\cite{disalvo2014making,jylha2013matkahupi,kuznetsov2011red}. In this work, we explore \textit{augmented reality} as another such ubiquitous technology, which introduces the layering of digital media onto physical spaces as a new affordance to hybrid activism. By investigating people's decisions and experiences around AR activism, we seek to understand how this new kind of activism fits into the broader activism hierarchy.

\subsection{Augmented Reality: Digital Activism in the Physical World}
For purposes of this paper, we define AR as a technology that embeds digital information, such as virtual images and audio, in physical environments. AR has evolved over the years to become a widespread consumer technology, with popular mobile phones now featuring AR capabilities~\cite{Schmalstieg2008AugmentedR2}. The wide reach of AR has opened up possibilities for a range of applications, including AR for education~\cite{wu2013current}, collaboration~\cite{piumsomboon2019shoulder,guo2019blocks}, health~\cite{mostajeran2020augmented,aoyama2020homemodar}, and, perhaps most commonly, entertainment~\cite{paavilainen2017pokemon}.

More recently, AR has shown potential for contributing to broad cultural shifts. In addition to individual groups that focus on direct action, incubators and collectives offer fundraising, technical support, and mentorship directed at AR-led structural interventions (such as overturning physical monuments) \cite{freeman2012manifestar, farber2019monument, fisher2021augmented}. Skwarek suggests that AR can provide an accessible, low-cost yet high-impact way to achieve activist goals -- particularly, spreading messages and sparking conversation around a relevant physical location or object. For example, \#arOCCUPYWALLSTREET was a global protest at the New York Stock Exchange and Wall Street area. Since physical protests were prohibited, activists held tablets using an AR application to overlay the area with images and audio submitted from global protesters. AR on relatively inexpensive devices thus helped to lower the (literal) physical barrier of protesting, while opening global participation for those limited by travel costs. In addition to protests, AR has been used to ``hack'' buildings, corporate logos, monuments, nature, and more with subversive messages~\cite{skwarek2018augmented,10.1145/3170427.3186526}. 

AR's immersive quality could also benefit activism. People can use AR to tell stories in more expressive ways, transporting audiences into a new reality defined by the creator. Immersive storytelling in mixed reality more broadly can arouse strong emotional reactions, including for those with whom people typically struggle to empathize~\cite{kors2016breathtaking,herrera2018building}. AR, being physically tied to the real-world, could build on these aspects by using stories to intervene on and at specific objects and places. In particular, recent work suggests the potential for AR to support \textit{placemaking}, i.e., the planning, design, and management of public spaces. For example, Sweetgrass AR (Figure 1C) is a place-based experience where college students and Indigenous knowledge keepers collaborated to create experiences that educate the settler population about the relationship between Indigenous and the colonized land. At specific locations of interest, they augment granite sculptures that carry strong symbolism to tell stories about Canada's Saddle Lake Cree Nation~\cite{mcmahon2019sweetgrass}. Through overlaid digital content, citizens, especially those from marginalized groups, have a venue to voice their opinions, discuss and critique decisions, and even reclaim their identities and histories connected to specific places~\cite{sanaeipoor2020smart,gonsalves2021radical,fisher2021augmented,almond2018we}.

Despite the potential promise of AR activism, we still have a limited understanding of how people are actually using AR for activism. What drives people to use AR for activism, and what are their subsequent experiences? To what extent do AR creators feel they are able to achieve their activism goals? As AR is an increasingly ubiquitous and evolving technology, we must understand both the opportunities and challenges of using AR for social change. To this end, we conducted an interview study with people who have created AR experiences for activism in order to explore the following research questions:

\begin{quote}
\textbf{RQ1:} Why do creators use AR for activism? \\
\textbf{RQ2:} What characteristics of AR do creators find conducive or detrimental to activism? \\
\textbf{RQ3:} What role does AR play in achieving creator's goals?
\end{quote}

\section{Methods}

\subsection{Pilot Study}

To ground our interview guide, we ran a pilot interview study with an initial set of eight activists who use AR and AR-adjacent technology such as VR. Focusing on U.S.-based creators, the pilot study examined people’s motivations and understandings of AR platforms across varied sites of public engagement, from a celebration of a historical march to community-based youth education initiatives in a historically Black neighborhood. 

Across creators' reflections on their varied contexts, AR technology emerged as a tool with the potential to raise awareness and spark radical engagements with everyday life. A Chicago-based artist spoke of the contrast between reading the news about gun violence and ``making it real, truly real'' through AR: \emph{``the fact that, my God, I am on the street right now. And there are young people who are on the street and they're thinking about, you know, what might happen to my life [now that] I lost my friend the other day on the streets.'' A Seattle-based artist emphasized the capacity for AR to make the impossible possible, a process with profound implications for Black people in a world rife with anti-Black racism. She created an installation that swapped the sky and the ground for Black and queer participants, looking to AR's potential for stretching physical limitations and visualizing aspects of our world that seem un-fungible: \emph{``the Black community could imagine itself out of the limitations of our physical reality  \ldots it's a radical imagination or reimagining of blackness and present that does not even imagine us to be here.''} Ultimately the interviewees viewed AR as a tool for prototyping the realities they wanted to build together, and less as a solution to societal problems. \emph{``Technology is not a savior in this situation,''} an interviewee explained. \emph{``I don't think it's technology that makes a difference, I think that it's community that does. And [AR] technology can be used to build community.''}}

Based on these responses, we refined our interview guide to dig deeper into people’s reflection on their experiences creating with AR technology. In particular, we focused on their hopes, expectations, and perceived outcomes for what they could accomplish with it, and the trade-offs they faced while using particular platforms.
We also used the pilot to explore possible limitations to remote interviews during the pandemic and to inform any shifts in our recruitment approach, such as focusing more specifically on AR creators. We describe the subsequent participants  and interview study protocol in the sections that follow.

\subsection{Participants}
We recruited participants involved in projects that created AR experiences for activism, based on the definitions of ``AR'' and ``social activism'' stated in the previous section. We initially found participants through a web search for relevant projects (e.g., using keywords such as ``augmented reality activism'') and referral from other participants. We aimed to include a wide variety of participants who engaged in activism projects that ranged across location, scale, and purpose. Of the 48 people we contacted, 20 people living across six countries agreed to an interview. These 20 participants worked on projects for a range of social causes, such as \changed{addressing institutional racism, environmental justice,} and more, with some projects initiated as early as 2009. Our participants are summarized in Table~\ref{tab:participants}. 

\begin{table*}[h]
\centering
\caption{Participant Table}
\renewcommand{\arraystretch}{1.1}
\label{tab:participants}
\small
\begin{tabular}{p{1.3cm}p{4cm}p{7cm}p{1.3cm}}
\toprule
\textbf{Participant} &
\textbf{\changed{Project Goal}} &
\textbf{\changed{Augmentation}} &
\textbf{Country}\\
\midrule

P1 \& P2 & Combat political corruption & Overlays text on political campaign advertisements & Germany\\

P3 & Expose unethical art donation sources & Overlays textual data, audio, and video stories on museum art & USA\\

P4 & Promote free speech & Displays cloud-shaped political text in the sky & USA\\

P5 &  Satirize US foreign relations & Distorts the visual appearance of US dollar bills & China\\

P6 & Pro-democracy & Juxtaposes historical news images and videos in an art installation & USA\\

P7 & Counter political misinformation & Overlays textual facts and data on political campaign banners & Germany\\

P8  & Sexual violence prevention & Overlays text and audio on a physical memorial & USA\\

P9 & Art democratization & Overlays digital guerilla art over physical paintings in a museum & Netherlands \\

P10 & Spotlight Asian American identity & Overlays digital version of a cultural painting over an iconic location & USA\\

P11 & Address institutional racism & Displays holographic historic images and objects at specific landmarks as part of a tour & USA\\

P12 & Promote diversity & Overlays interactive narrative text and animations on one's surroundings as part an educational game; displays holographic statues at specific locations & USA \\

P13 & Celebrate women of color& Displays an artistic holographic monument consisting of images, audio, and animations at a specific location & USA\\

P14 & Black history awareness & Overlays videos over specific landmarks as part of a tour & USA \\

P15 & Indigenous culture awareness & Overlays videos on cultural sculptures to tell stories & Canada \\

P16 & Celebrate women of color & Displays artistic holographic monument featuring visual design and audio at a specific location & USA \\

P17 & Environmental justice and awareness for marginalized groups & Displays holographic objects and avatars in one's surroundings, with which one must physically interact as part of a narrative game & USA\\

P18 & Celebrate women of color & Displays a holographic avatar narrating historical stories at specific locations & USA \\

P19 & Black history awareness & Displays educational holographic sculptures in one's surroundings & UK  \\

P20 & Black history awareness & Displays holographic avatar guiding a scavenger hunt and city tour & USA \\

\bottomrule
\end{tabular}
\end{table*}

\subsection{Procedure}

We conducted 60-minute semi-structured interviews with participants over Google Meet. All participants were interviewed individually, with the exception of P1 and P2, who had jointly worked on a project and asked to be interviewed together. 
Building on our pilot study, our interview protocol included questions about their work background, their understanding of AR activism, their rationale for using AR, the influence of AR on their project, and their broader experiences with AR. In exchange for their time, we offered participants a \$50 gift card or a \$50 donation to a non-profit organization\footnote{Participants chose between the American Civil Liberties Union, Habitat for Humanity, and the National Association for the Advancement of Colored People Legal Defense and Educational Fund}.

\subsection{Analysis}
We analyzed transcripts of the interview audio files using 
an inductive analysis approach~\cite{thomas2006general}.
We used the ATLAS.ti qualitative analysis software to code the transcripts. We first generated inductive codes using a subset of transcripts, looking for similarities across participants' experiences and labeling the ideas that emerged (e.g., ``making the invisible visible''). We conducted three rounds of open coding on three transcripts each. We created a codebook from the generated open codes, focusing on codes that were prominent across interview transcripts and relevant to our research questions. Two independent coders validated this resulting codebook on another subset of transcripts, discussing any disagreements to ensure high inter-rater reliability. After achieving Krippendorff's $\alpha$ above 0.8, the two coders divided and coded the rest of the transcripts. We then grouped similar codes together to form higher level categories, which we further discussed and refined to identify cross-cutting themes around perceptions and use of AR activism. Throughout this process, we revisited the interview transcripts when necessary to clarify codes. We organize these themes in the following section according to our research questions around people's decisions to use AR for activism, the characteristics of AR activism, and the impact of AR activism.

\section{Results}

\subsection{Why do creators use AR for activism?}
We wanted to understand the factors that influenced participants' decisions to use AR for their projects. These factors included prior AR-relevant experience and the perception that AR content is more engaging and safer to deploy, given today's undefined legal and material constraints. At the same time, participants expressed concerns around the potential for future legal or ethical repercussions.

\subsubsection{Prior AR-relevant experience}
Fourteen of our participants shared that their paths to using AR for activism projects included their prior experience with AR and/or development of AR-relevant skills such as programming, design, and storytelling. 
P3, a professor of design and computing, outlined their path to using AR for activism through developing myriad relevant skills:
\begin{quotation}
\emph{``My current work explores \ldots a lot with AR [and] experimentation and challenging technocratic power through \ldots creative experimentation and hacking. 
I come, really, to AR from starting with exploring things like cryptography, steganography, obfuscation \ldots which lead to looking at multilayered narrative. How can we augment the physical space with the virtual space, and create layers of narrative?''} - P3
\end{quotation}

P3's years of exploration in both emerging technologies, including early versions of AR, and steganography (concealing information in non-sensitive data, e.g., embedding a message in an image) motivated their use of AR and enabled them to create multilayered narratives around ``dirty money'' in the art world. This experience manifested in how their project presented previously obfuscated information to museum-goers, using AR to overlay visual statistics and audio recordings about the opiod epidemic on top of museum art bought with opiod money.

Other participants shared diverse experiences influencing their use of AR that were not necessarily technology-centric. P13, a historian, artist, educator, and cultural organizer whose project augmented locations with digital artifacts about a pioneering Black woman entrepreneur and philanthropist, shared: 

\begin{quotation}
\emph{``I call myself an unbound artist because I'm not tied to any one medium. I usually center my work on a lot of deep research in the world, but then also research in the books and in a lot of archival research. My work is really about \ldots how do I take the knowledge that has been ignored from dominant narratives and create a space for more people to see it outside of the academic towers. I'm a memory worker \ldots specifically as it relates to the work that I do with monuments.''} - P13
\end{quotation}

As a ``memory worker'' who aims to share overlooked histories with wide audiences, P13 was drawn to AR as a means to combine their skills in history, performance art, and multimedia, using archival image and audio data to tell stories about a historical figure and imbue physical locations with new narratives.

As seen in these two exemplars, participants drew from their background to envision rich multilayered narratives for their activism projects. Their prior experiences motivated their use of AR as a means to compose those narratives through a combination of skills in technology, history, art, storytelling, and so on.

\subsubsection{Ease of creating engaging content}\label{easeofcreation}
Participants described experiencing a relatively low barrier to creating compelling experiences for activism in AR using modern technology.
Twelve participants had worked with AR before, but described requiring fewer resources to create AR experiences for activism compared to their experiences in the past. 
P10, a multimedia artist who created an AR installation centered on \changed{Asian American} identity, described changes in AR development technology in their experience:

\begin{quotation}
\emph{``I think, nowadays, the platform is much [more] intuitive for artists to adapt. There's not much coding. It's much easier, but at the beginning, it was a challenge.''} - P10
\end{quotation}

Here, P10 speaks to how modern AR creation programs have introduced UX features and supporting material (such as prefabricated digital assets and tutorials) that make it easier to create AR content, even without programming expertise.

Building on this insight, many participants felt that it was easy to use AR platforms to create new, compelling interactions that could enhance how they deliver their message, especially when tied to the physical world. For example, P15 described the ease of creating an AR experience that layered videos telling stories about Indigenous heritage onto physical sculptures important to cultural preservation:
\begin{quotation}
\emph{``We didn't use animation or anything, it was essentially just an interface where you could click on videos, and then the video would play but just layered videos, it's pretty simple AR. We chose HP Reveal, [which was] YouTube-style, where you just upload a simple video.''} - P15
\end{quotation}

P15 felt the AR tools they used enabled them to create a simple yet rich experience, with which they aimed to create new understandings and perspectives around the history of Indigenous communities.

Taken together, participants felt AR is an approachable yet powerful medium for activism, especially with the rise of more novice and non-programmer friendly platforms and features.
The ability to create immersive content without coding, including the ease of uploading videos and attaching them to physical objects and locations, made AR an attractive tool for delivering their message without specialized skills or expensive resources.

\subsubsection{Potential legal and ethical consequences are still being shaped}
Eleven participants noted that AR activism seems less risky than traditional activism, even as the consequences of those risks specific to AR remain uncertain and under-defined. P1 and P2, artists who collaborate on global activist interventions, described how AR enabled them to accomplish what would be too risky if attempted through another medium. They remotely created an AR experience to expose fraud by  augmenting the election campaign ads of a corrupt government.  They noted, 

\begin{quotation}
\emph{``If you would have painted the words on the posters, I can guarantee you these people would be imprisoned, if not dead by now \ldots ''} - P1 \& P2
\end{quotation}

By augmenting posters, rather than physically painting over them, these participants were able to distribute their message, tied to physically relevant objects, while avoiding physical risk. 
With AR content existing only on a digital rather than physical layer, creating controversial content feels less high-stakes for activists, especially compared to the risks of engaging in gladiatorial forms of traditional activism outside of AR that P1 and P2 mentioned.

At the same time, half of the participants raised concerns over the lack of certainty about legal ramifications associated with activism in AR, even if it enabled their work. P7, an innovation art director who created an AR project that commandeered Trump election signs, spoke to these uncertainties: 
\begin{quotation}
\emph{``I was just the whole time, a bit afraid \ldots because the Trump campaign is known to sue anyone and everyone, and I was a bit afraid of, `Where do we overstep? Can we use this name? Can we use the look of his sign?' \ldots we also brought [a law student friend] in to basically double-check everything just to make sure like what are the legalities and everything of this project.''} - P7
\end{quotation}

P7 highlights that while activism in AR can mitigate risks associated with traditional forms of activism, it still brings its own set of potential negative consequences.
The lack of legal precedent on controversial uses of AR does not eliminate the risk of liability and prosecution, especially when AR content involves material typically subject to legal protection outside of mixed reality, such as US presidential election campaign materials.

Alongside the immediate risk of physical harm and legal liability, creators worried about long-term ethical repercussions. In particular, creators struggled to make sense of their use of platforms that are created, sanctioned, or funded by companies with strong ties to ``anti-social'' activity. P15, a settler scholar working with Indigenous communities on technology development, described the complex ethical implications of creating accessible digital media based on Indigenous knowledge: 
\begin{quotation}
\emph{``I think for the communities I work with, they need a balance between user friendliness, which is very important, [and] low costs, hosting fees, design fees \ldots I think the biggest thing is that control and ownership \ldots this Indigenous data sovereignty is really, really strong now and I think for Indigenous peoples in particular, where is their data going? How can you access it? \ldots They have a lot of protocols about even what time of the year you can access information and stuff like this.
If there was some off-the-shelf platform, maybe free and open-source that would allow for local hosting of data, that kind of thing, non-commercial, I think [they very much] don't want to commercialize it unless somehow it benefits the community.''} - P15
\end{quotation} 

For P15 and their team, one of the main criteria for considering AR and choosing an AR authoring tool relates to the level of control they felt they had over their own data. Deciding what data to retain and how, who drives changes to data practices, and who is responsible for ensuring equitable access was a bigger constraint than any other technical or logistical barrier. 

From well-established democracies to countries governed by authoritarian regimes, AR activism can appear to lessen risks typical for traditional activism, though with murky legal and ethical bounds. 
On the one hand, participants point out that AR-based activism can remove the need to physically alter the world or even be physically present to augment it, which can reduce associated risks \changed{and make the use of AR more appealing}. On the other hand, due to the emergent nature of AR activism and a reliance on proprietary and commercial platforms, participants described \changed{some hesitance in using AR due to} the lack of understanding around legal risks and social harms associated with fewer (or selective)  restrictions.

\subsection{What characteristics of AR do creators find conducive or detrimental to AR activism?}
Participants discussed features of AR that facilitated or limited their activism projects.
This included characteristics of the technology itself, as well as the development platforms they used and collaborations they formed to create their projects.
In particular, participants felt that the ability to visually layer narratives in AR could be beneficial for changing perceptions and reshaping physical places. 
At the same time, despite the aforementioned lower barriers for creating AR content, participants still faced some challenges when using modern AR development platforms for their projects.

\subsubsection{Potential for changing audience perceptions}
Like many activism projects outside of AR, 15 of our 20 participants aimed to change people's perceptions through their AR activism work. Participants described how AR could expose audiences to crucial, yet under-recognized, aspects of their social cause through immersive narratives.
P20, an artist and professor of new media who created a narrative AR scavenger hunt centered on African American history, shared how they hoped AR could change perceptions of Black and Brown women:
\begin{quotation}
\emph{``[As] someone who'd like to become a parent, I'd like the landscape to look different even if it's through augmented reality for a young person so that they can get a vague sense that there's someone who's really invested in letting them know that they come out of a legacy of resistance, of activism, of agency, of making significant change. A legacy of people making significant changes and alterations to the would we live in today. This world did not come into existence without Black and Brown people and women, it was made by them.''} - P20
\end{quotation}

Using AR to overlay videos of reenacted stories about the hidden histories at specific murals, plaques, and other physical urban spaces, P20 designed the scavenger hunt to give people digital windows into the past that could change their existing perceptions about the city. In particular, they aimed to make the invisible presence, agency, and work of marginalized individuals, in this case African Americans, \textit{visible} in the world.

In addition to making the invisible visible, participants described a sense of \textit{immersion} in AR, deeply connected to the multilayered narratives they created, that could further influence people's perceptions.
P14, a history professor and author, worked on a narrated AR tour of a historical park in the US where African Americans flourished during the Reconstruction Era. Though their project was publicly available, they discussed immersion in the context of students:

\begin{quotation}
\emph{``[I think] this additional layer of reality, it brings the ability to immerse people in stories. I think that sort of immersion is what's important. What I see happening is a good thing for me because I see increasing abilities of technology to allow me as a teacher to immerse my students in the story. By immersing themselves into the story, they become part of the story, the story becomes personal. It is not just something that happened to somebody else 200 years ago. It's something that, in a way, is sort of happening to them. Everything seems to be self-centered these days. If you can relate to the story and that helps people learn it, then I think that's great.''} - P14
\end{quotation}

P14 hoped that audiences would feel immersed in AR's ``layers of reality,'' and therefore more connected to the historical content as narratives relevant to themselves. They aimed to establish this connection in their AR tour by layering aural and visual historical content (e.g., music, photographs) over key physical areas, such that people could feel transported to the past where they stood.

Overall, \changed{participants felt that the immersive characteristic of AR, which ties narratives to the physical world, could help their audiences} better understand new perspectives and connect with their message, such as the importance of forgotten moments and people of history.

\subsubsection{Reshaping the physical world}
Beyond surfacing invisible information to change perceptions, 17 out of 20 participants also discussed using AR to \textit{reshape} elements of the physical world, such as public spaces and objects, by using AR to envision how they wanted the world to look. P8, for example, contributed to a project that aimed to reshape a university space associated with sexual assault:

\begin{quotation}
\emph{``We started brainstorming different ways to use AR like a really lightweight means for someone to take a fairly ubiquitous technology and be able to use it to basically modify the space in ways that they wanted to without needing the prerequisite political or physical power to be able to modify that space.''} - P8
\end{quotation}

P8 highlights how the virtual nature of AR can facilitate reshaping a physical space without having to grapple with the typical requirements for doing so.
In their project, they used AR to overcome the hurdle of campus administration, which had installed a memorial for a sexual assault survivor but rejected quotes the survivor proposed for a plaque. Through AR, they could virtually bring the plaque with the desired quote read in the voice of the survivor to campus, along with voiced-over letters from other students.
This layering enabled them to reshape the memorial, such that viewers could better understand its context and meaningfulness.

Similarly, P11, a student who created an AR tour on their university campus around legacies of slavery and racism, spoke about using AR to reshape a part of their campus museum: 

\begin{quotation}
\emph{``When we did the stop [on the tour] for the museum where we studied the [a physician who] collected hundreds of skulls, and organized them by race, and used different measurements to create a racial hierarchy, and medical misinformation that's still extant in the medical field today. We did a kind of dome. Not only are you inside of a skull, but you're also being encompassed by all this information. You're being surrounded and immersed by the institution, by the original things that founded white supremacy, much like white supremacy surrounds us as a country.''} - P11
\end{quotation}

By manifesting a large virtual skull that people could walk into, this participant reshaped the museum on their AR tour to reflect the hidden structures in the country's and educational institutions' histories.
Distinct from using stories to reveal or recreate the past, P11 created a new artifact as symbolic representation of the unethical legacies that the university physically perpetuates on campus. \changed{Thus, participants felt that AR's virtual nature, unbound by involved processes for changing physical space, could help audiences envision and contextualize spaces in new ways.}

\subsubsection{Interdisciplinary collaboration and support} 
As most participants mentioned, a diversity of backgrounds and skillsets are relevant to crafting activism projects in AR that are high quality and effective experiences. 
Seventeen participants reflected that AR activism more often than not requires interdisciplinary collaboration between developers, designers, storytellers, etc. More specifically, participants described collaborating with independent art groups (P1, P2, P4, P5, P9, P10, P13, P18, and P19), non-profit organizations (P6, P19, and P20), academic institutions (P3, P5, P8, P9, P11, P14, P15, P17, and P20), and private companies (P12, P13, P16, P18, and P19).

P18, a film and media maker whose project overlays a city with 3D animated renditions of historical figures and buildings to retell the lived experiences of a civil rights activist, described a large collaboration network that facilitated her work. 
An initial tech fellowship and residency program, as well as starting her own production company, helped her form connections with technology companies, the film industry, and artistic partners.
Her project collaborations included a local 3D modeling team, a local historical society to ensure historically accuracy of AR content, an award-winning actress to record voiceovers and motion capture for animations, and a local AR company to help her put everything together. 
P18 described how collaborating across disciplines supported the quality of her AR activism project:

\begin{quotation}
\emph{``Yes, I had companies that \ldots would take my design and bring it to life \ldots then having other local companies do prototypes for me so that I could see how [the main historical figure] is moving. [I would ask them] `Does the motion capture that I got really work? Can they add some animation? Can they add some a flare here, a flare there, some special effects that enhance where people are looking and directing their eyes, and getting them to go to the next step?'''} - P18 
\end{quotation}

P18's interdisciplinary collaborators all contributed to bringing P18's vision for the project to life, and addressed many challenges she may not have been able to tackle on her own or had limited expertise in.

However, interdisciplinary collaboration for such projects can have negative aspects. 
While P18 appreciated her collaborations, she also discussed difficulties in the collaboration dynamics:

\begin{quotation}
\emph{``Again, these are things that the ebbs and flow of like, do you have enough money? Do you lose your space? [Beyond that] I get paired with these very nice Caucasian companies \ldots these white boys they're not interested \ldots  I got a bunch of work to do to add a layer of, `Now I got to convince you that this is important or that maybe you should stay up another hour to help me get this done?' 
[The project's main character] doesn't strike them as cool, `She's not cool. She's not that interesting. Oh, she's not a superhero. Oh, she doesn't have big breasts. She's not a sexualized object.' That's the stuff that makes me feel like, `Oh, I wish I had just not been paired with certain companies.' They're wasting my time and they're wasting everybody's time.''} - P18
\end{quotation}

Here, P18 speaks to how resources like money affect interdisciplinary collaboration as well as how racism and sexism, in the face of her intersectional identity as a Black woman, affects how collaborators treat her. 
These collaboration issues can become obstacles in pursuing an AR activism project, as well as more broadly affecting the creator's quality of experience. 
Thus, while many participants reflected on interdisciplinary collaboration being conducive to their project's success (including those who did not have collaborations but wished they did), P18 illuminates the importance of the collaboration being founded on a professional, healthy, and respectful relationship.

\subsubsection{Uncertainties and difficulties in AR development and distribution}
Seventeen participants struggled to keep up with shifts in AR development platforms.  AR creators reported using a variety of development platforms, including ARCore, Artivive, HP Reveal, LayAR, Lens Studio, Metaio, Reality Composer, Unity3D, Vuforia, WebAR, and Zappar.
P4, a multimedia artist whose AR work displayed political messages in the sky, discussed how the varying lifespans of AR platforms affected their work. 
They were particularly frustrated with the disappearance of LayAR, their previous platform of choice:

\begin{quotation}
\emph{``[LayAR was] probably the most advanced at the moment. All of a sudden, they're gone. All right. I had a few projects I was using on that platform too \ldots ''} - P4\end{quotation} 

Other participants praised and trusted this platform, and found that its abrupt end affected not only the continued development of their work but also its distribution, as people could only experience the content on this platform. 
Upon losing a valued tool like LayAR, creators, especially those without programming expertise or who have built platform-specific expertise, must spend time and energy learning how to effectively use another platform. 
Moreover, translating their work to new platforms was not always straightforward, as platforms may not support the same features, e.g. uploading videos YouTube-style.

Despite improvements in modern AR creation tools for non-developers (e.g., interfaces to easily attach AR content to the physical world discussed in \ref{easeofcreation}), eight participants still described difficulties related to specialized skills aside from coding.
P9, who has experience with various AR development platforms, described how working with those platforms can be resource-consuming due to platform-specific development and publishing requirements.
However, P9 also noted that some platforms are considerably easier to get started in and efficient to use, such as when comparing Unity and Lens Studio:

\begin{quotation}
\emph{``With Unity, I often end up working a lot of time with \ldots things not working \ldots There's so much `not working' in Unity. If you're a beginner, it's difficult \ldots  [In Lens Studio] you can't make it too complicated, so it saves you a lot of time. If I'm collaborating with somebody in a Unity project, they keep telling me about things I've never heard of and the good thing with Snap is that I think I've seen most of the tools at hand now, the mechanisms, and that allows me to be an expert quite rapidly. With Unity, that's not the case, and when you're an expert with the tool you can make anything.''} - P9
\end{quotation}

P9 also compared the publishing approach on both platforms:
\begin{quotation}
\emph{``With the Snap Lenses, you can say, `Well, [the project is] there, and if you want to use it just download the app and there you go.' A lot of the technical struggles are taken care of by such a platform tool \ldots it's a WYSIWYG tool \ldots. With Unity, you always have to think, `Okay, what are the platform-specifics' and then clients want high-end renderings and they want baked lighting and all this terrible stuff that takes a little time away from the essence of making something interactive, something cool, something new.''} - P9.
\end{quotation}

This comparison brings attention to characteristics of AR development platforms that can facilitate or hinder the creation of activism projects in AR.
In particular, creators saw ``what you see is what you get'' tools, mechanisms, and publishing processes as enabling efficient completion of projects as well as simplified sharing. Unreliable and unnecessarily complex features and publishing requirements can detract from building platform expertise and actually creating content.

At the same time, it is important to note that platforms may be designed for specific purposes, to which participants had to adapt for their goals of activism. For example, Lens Studio is a platform for developing AR filters for Snapchat, while Unity is a game engine for creating AR \textit{in addition to} VR, 3D, and 2D games that can be published to a variety of distribution platforms.
Thus, certain development platforms may be more or less conducive than others to creators' goals (barring unforeseen discontinuations) depending on creators' experience levels, project complexity and longevity, and target distribution.

\subsection{What role can AR play in achieving creators' AR activism goals?}
Participants reflected on their projects' impacts, and how AR may or may not have contributed to achieving their activism goals.
It is important to note that we discuss impact based on participants' \textit{perceptions}, including their aspirations through their work, and what they heard and observed from people who engaged with their project.
Participants shared that while they feel AR has the potential to influence physical action and strong emotional reactions, AR also had limited reach to audiences in terms of awareness of the project and ability to access and use it as intended.

\subsubsection{\changed{Eliciting} physical action and strong emotional reactions}
Fourteen participants discussed the potential of AR to affect physical action and/or strong emotional reactions among their projects' target audiences. P8, the aforementioned participant who introduced an interactive AR plaque for a sexual assault survivor at a university campus, shared how their project may have contributed to the university's decision to install a physical plaque at the memorial:

\begin{quotation}
\emph{``What is interesting about the plaque [that the university physically installed] is that it looks like the [AR] plaque that we designed for our project. [Our AR project] ended up being an inspiration to what actually ended up happening in the physical space \ldots .''} -P8
\end{quotation}

P8 further noted that while they did not issue any official statement with the university, their team did discuss their project with the university administrators, including its context and the barriers around installing a physical plaque. They felt that this discussion, along with the virtual plaque they created, may have contributed to the ultimate outcome of having a physical plaque installed.

Participants perceived that AR could also elicit strong emotional reactions in audiences. Because AR is a novel technology, seeing and hearing an additional layer of reality can be a completely new and emotional experience for many.
P5, a professor of multimedia filmmaking who used AR to alter US currency to highlight foreign relations, stated: 
\begin{quotation}
\emph{``Technology is quite emotional for people, our world being shaken up by what's possible for you to see or hear combinations that you can make in your immediate perception \ldots it's maybe this positive thing and so we're still taking this traumatic moment of things being shaken up for us and the possibilities and what you thought was possible in the world being changed, but instead of just responding with this fear, instead, we have this strange experience of fascination, of engagement, of small joy.''} - P5
\end{quotation}

P5 describes a strong emotional experience that they believe someone can have while engaging with AR, which could make the user more receptive to content that challenges what they previously thought or believed. Participants further note that AR's novelty alone cannot achieve their activism goals, but paired with well-crafted content, such as the compelling multilayered narratives participants aimed to integrate, they may produce something emotionally impactful.

Relatedly, one participant highlighted the unique role they believed AR played in sparking emotions, based on the experience of someone who interacted with their project.
P17, an artist, activist, and academic, created a project based on their own lived experience that exposed how global warming is affecting the lives of the already most vulnerable. They compared the use of AR in the project to their previous work on the same topic:

\begin{quotation}
\emph{``I didn't want people to look away from the world, I wanted people to look at the world, and I thought that augmented reality was a way to do that \ldots What I want is for people to think about [climate change] and engage with it and look at it and face it and face the fact that we have to do something \ldots. [Compared to my previous web-based game on climate change], I think that the AR project has a more emotional impact\ldots[someone who interacted with it] said to me that they felt implicated because they had to move their body to engage with the work. As opposed to the disconnect of sitting in front of a screen and just clicking on a mouse, which is such a familiar feeling that it can be really disembodying, dissociating, disconnecting for people I think.''} -P17
\end{quotation}

P17's project required users to physically move around, collecting oxygen capsules and narrated stories about climate change's effects while navigating an AR forest. Drawing from their prior non-AR work and what they learned from users, the participant felt that the embodied interaction in AR could emotionally immerse audiences in the content without virtually removing them nor their gaze from the physical world, unlike other forms of technology that inherently draw attention away from the physical world.

Additionally, a few participants discussed the potential for AR to elicit strong emotional reactions in marginalized individuals in particular. P17, for instance, described another interaction they had with someone who engaged with their project, which focused on the effects of climate change on marginalized groups: 

\begin{quotation}
\emph{``[She said] she felt just very good [about] having felt represented \ldots [she said] as somebody living with chronic respiratory illness, she felt like nobody had made artwork about having asthma and being in the middle of a pandemic or being in the middle of a climate change event. [My project's goal] was to point out how climate change is already harming people and disproportionately harming marginalized people, like trans people and chronically ill people and immigrants. That felt like a real success to me, that somebody from one of those groups said like, `Oh, I feel seen by this work.'''} - P17
\end{quotation}

In addition to emotionally engaging people with a social cause, participants like P17 suggest that AR may have the potential to empower marginalized individuals. They see AR as intentionally centering under-recognized narratives and voices in a layer of reality that blends with the physical world. One participant, an artist, creative technologist, and educator who created monuments of Black and Brown historical figures, further envisioned how such empowerment could have implications for the future:  

\begin{quotation}
\emph{``Part of the work that we're doing is \ldots being able to inspire the kids to do their work \ldots in the future, I see us \ldots allow[ing] students to design their own monuments or design their own experiences [in AR]. I think just allowing people to see that they can be more than just a consumer of these technologies, but they can actually build them, and scale them up, and be able to create their own tech companies or whatever, art companies, [etc.]. That's a big part of [our  AR activism project] as well, and why we're always speaking and talking, and why we're focusing on the kids now.''} - P12
\end{quotation}

Participants thus aspire for AR to not only help people feel seen and uplifted, but also help people learn about and celebrate the legacies of those who look like them as sources of inspiration and affirmation of their own capabilities.

Overall, participants believed that AR could help achieve their activism goals. For creators aiming to reshape physical spaces, AR \changed{could play the role of} raising awareness and prompting people to consider changing the spaces re-envisioned through AR. By grounding identity-specific AR content in the physical world, creators hoped AR could spark highly emotional reactions to their projects.

\subsubsection{\changed{Limited reach}}
While participants saw potential in the impact of their projects, seventeen participants felt that AR activism still has a narrow reach. 
In particular, they felt that AR is not quite a ubiquitous technology yet, which can limit people's ability to understand and interact with it.
The aforementioned P12 
shared:

\begin{quotation}
\emph{``I was showing AR \ldots to seniors. It didn't work very well. They didn't understand it. I tried to explain it and they weren't getting it. I don't think anything I could have said was going to help them understand it. I feel there's a blocker if you're too old to understand what is going on in augmented reality on average. I'm not saying every person who's a senior can['t] understand augmented reality, but that is my experience.''} - P12\end{quotation}

Though creators may target a variety of audiences for their projects, not all audiences are necessarily familiar with AR or relevant technology. This can limit people's ability to understand their project and connect with the underlying message. Moreover, participants shared that they encountered people who did not know how to find these experiences in the first place. Some participants attempted to address this issue by using social media to advertise and instruct people how to access and experience their projects.
For instance, P19, a lawyer and community organizer who created AR educational materials centering Black people's historical achievements, described how posting on social media helped distribute their project to wider audiences including adults, children, and potential partners and supporters:

\begin{quotation}
\emph{``We just put it on social media and then a lot of people \ldots picked it up so that I think that also on social media, it looks really good. [A singer in the UK] put it up and she [had] over 300,000 followers \ldots I think social media was the main pusher. Then, after that, we had [a big company] approached us and a couple of these big tech companies [were interested in the project] \ldots ''} - P19
\end{quotation}

Participants describe challenges in people's general awareness and exposure to AR. While social media can be one tool to share their project to a large network, particularly if they have access to social media influencers, some populations may struggle with using and understanding their project even if they engaged with it. Participants highlight that the varying levels of access to AR and ability to engage with it can be a major barrier, as being able to spread their message and reach target audiences is key to their activism work.

In addition to limited engagement and awareness, seventeen participants noted that AR distribution platforms can also limit their reach. 
P17 described how art venues and distribution platforms can affect reach: 
\begin{quotation}
\emph{``I feel like very few people know that my app exists. I think it got some visibility from being [featured on an art venue] and winning an award. Mostly, people just experience my work through art venues \ldots that's okay with me, but it would be nice if a lot more people knew about the project. I put it on the [Apple] App Store. One challenge is [that it is] artwork that deals with difficult themes, so I labeled the app in the App Store as 17 and up. It's for adults, it deals with violence, just descriptions of violence. That may be a barrier to people finding it. I think another barrier is that the way I made [the project] platform-specific, so it's only for iOS.''} - P17\end{quotation} 

From art venues to the Apple App Store, AR activism projects may not gain exposure without being featured or winning an award, and may only circulate within specific, niche communities, like multimedia art galleries or museums. 
Because of limited methods of distribution, AR activism works may not be able to have the reach that creators aim for, which could negatively affect achieving their projects' goals.
It is important to note that even if creators find publishing on some platforms easy (e.g., Lens Studio, as previously discussed), they may still be limited by that platform's distribution capabilities (e.g., search tags and content restrictions, platform specificity, and requirements to download and register for specific apps).

Also potentially affecting reach is that much of AR content, including some participants' projects, can only be experienced at specific locations in the physical world.
Our participants discussed how this could limit exposure to audiences, given the need to travel to that location, but at the same time make the experience more special.
Specifically, P16, who created an AR monument tied to a public space in the physical world, reflected on this tension that surfaced when their project was shared by someone else on social media:

\begin{quotation}
\emph{''[They think,] `why do we need to tie [the project] to a location when there's no reason to do that?' In [their] mind, building an AR monument and then making people go see it when you have your device right here and I could see it on my desk just as easily, we're putting an artificial constraint on something that doesn't exist in this medium. I can totally understand that philosophically, except that I can also see the value in giving a specialness to something. \ldots what I struggled with about one of the qualities of AR is it can be anywhere all the time \ldots If we're talking about monuments and locations and specific histories, as a medium that's ever-present everywhere, does that cheapen it for me to be able to see it on my desk, or does that facilitate something for me and make it better? I don't know.''} - P16
\end{quotation}

P16 describes how rather than a constraint, location can be an important design decision or feature that contributes to AR activism goals. As described in the previous sections, many participants relied on location to immerse people in content central to where they stood. At the same time, people may not be able to access those locations, limiting the content's overall reach. Keeping these potential trade-offs in mind, the role of location-specific AR in creators' activism goals relies heavily on their specific 
activism goals, content, and target audiences.

Overall, participants felt that limited awareness of AR and their projects, varying audience readiness to engage with AR, distribution platform constraints, and possibly even location-specific features can negatively affect their projects' reach. Limited reach can in turn hinder their connection with audiences, and thus, their goals of changing their perceptions of the world. Still, as shared throughout our results, participants perceived AR as a powerful medium for activism due to its potential to facilitate the creation of compelling multilayered narratives that can elicit physical action and strong emotional reactions.

\changed{
Returning to our research questions, we learned of a variety of motivations, values, and expectations that drove creators' use of AR for activism, and their subsequent experiences. We found that creators used AR for activism due to relatively low barriers to creating compelling content -- technically, legally, and physically (RQ1). Creators highlighted that AR's ability to immerse people in narratives that reveal or imagine realities could be conducive to activism by changing perceptions and raising awareness. Conversely, creators discussed challenges that AR introduces which could be detrimental to activism, including limitations to creating, maintaining, and distributing content (RQ2). Finally, creators described their perceptions of AR's role in achieving their goals, specifically in influencing physical changes in the world and empowering their audiences. At the same time, they questioned their ability to reach large audiences given limited access or distribution of their projects (RQ3). In the following section, we discuss these findings in the context of prior work, and highlight open questions and opportunities for future research.}

\section{Discussion}

Across our interviews, participants saw AR as an emerging form of \textit{hybrid activism} \cite{heaney2014hybrid} that incorporates digital media in the physical world for social change. As creators, participants provided a variety of testimonials, from guerilla installations at museums exposing questionable art donations to geofenced memorials highlighting the voices of oppressed people. They perceived AR as an effective tool for expressing unique and engaging messages, with the potential to change perceptions and help people reimagine public spaces. But they also faced several challenges reaching broad and diverse audiences and noted emerging ethical and legal considerations related to user control, author responsibility, and platform circulation. Ultimately participants sought to use AR to elicit strong reactions and inspire action among their audiences within a fraught and precarious sociotechnical landscape.

Recalling George and Leidner's hierarchy of digital activism~\cite{george2019clicktivism}, our work suggests that project creators view AR activism as enabling a high level of engagement, potentially within the gladiatorial level. Like hacktivism, they see AR activism as requiring technical skills (e.g. programming knowledge) in order to take action. However, unlike hacktivism, creators see AR as becoming accessible to the general public and currently having few legal restrictions. Additionally, creators perceive AR as having the potential to generate high impact by triggering audiences' emotional reactions, which might catalyze civic participation. At the same time, creators worry that the platforms' limited reach may hinder the impact compared to hacktivism or other gladiatorial level actions.

\subsection{Open Questions and Opportunities}
Looking beyond AR's promise for social change, creators point to several open questions and opportunities. Below we summarize these lessons around themes of interdisciplinarity, platform control, discoverabiity, accountability, and radical change.

\subsubsection{Supporting Interdisciplinary AR Content Creation} To create immersive experiences that could elicit strong reactions, participants required specialized expertise and engagement with collaborators and organizations from multiple fields. Many participants established interdisciplinary collaborations to overcome technical limitations and generally create higher quality and more impactful projects. At the same time, collaboration can bring challenges in coordinating resources as well as perspectives around the topic and vision of the project. Worth noting, and illustrated by P18, these collaborations can also reproduce unethical patterns (such as racism and sexism) that are embedded in broader social contexts which the technology has limited (or no) means to directly address. 

Informed by these experiences, we encourage researchers, activists, and supporting organizations to reflect on the conditions of these interdisciplinary collaborations and the values that underlie these interactions. Within research and activist organizations, this work involves adapting community-driven strategies that emphasize long-term engagement, participatory planning, and the placement of technical decision makers in service of existing practices \cite{creativereact, boyd2013beautiful,costanza2020design}. For those working specifically on creating AR authoring tools, we encourage expanding support for the integration of well-established standards (e.g. widely used 3D models and video formats), mechanisms for collaboration (e.g. intuitive data management systems, discussion groups, and brainstorming boards), and greater number of practices and skill sets (e.g. traditional painting, physical installations, architecture, and informal education) that are already familiar to a range of potential users. Inspired by efforts like Mozilla's WebXR \cite{maclntyre2018thoughts}, AR development platforms should also consider the integration of universal and public AR project formats to facilitate content access and the migration of projects among platforms, so future activists can have more reliability and flexibility for their work.

\subsubsection{Confronting Maintenance and Proprietary Barriers}
Participants saw the development of impactful AR activism experiences as just the first step. Although AR offered an entry into civic engagement, they described struggling with the maintenance and distribution of the projects. Sometimes this concern reflected uncertainty around the future availability of their projects on particular AR platforms, which could shut down at any moment. Other times participants reflected on the consequences of not having full control over the content and experiences developed using proprietary platforms. In addition to losing access to platforms, and thus losing time and effort, their remarks point to the importance of platform ownership itself. When creators like P15 centered community, they sometimes viewed a platform's capitalistic values as running counter to their own goals. They expressed feelings of unease about the corporate partner managing their platform and that decides what could be seen, extracted, recirculated, and more. Their concerns focused on the potential for surveillance, monetization, and various harmful structural interventions that lay outside their control.  Even compared to social media platforms like Facebook, where organizers can share web links that make content accessible outside the platforms, AR currently tethers organizers to particular corporate or commercial interests. Participants' reflections as creators suggest the degree of control they have over the AR platform — how the data gets used, where/how it gets circulated, and to/for whom — is core to understanding AR's long term potential for building conditions for social change.

\subsubsection{Improving AR Discoverability}
Participants suggested that the distribution of an AR activism project can be as challenging as the creation of the project itself. Their reflections suggest that improving the discoverability of AR content would go a long way toward improving AR access and broadening its impact and reach. Indeed, discoverability is a known issue with AR technology, where people may not know about existing applications or realize they are available, and creators may not actively promote or have the means to promote their content~\cite{de2013mobile}. While work in e-commerce has highlighted the need for better calls-to-action while browsing and better transitions into AR ~\cite{harley_2020}, increasing discoverability more broadly might involve nurturing open source and other grassroots AR tool-kits that enable more flexibility and adaptability around access conditions. Conversely, it could involve further integrating AR authoring and viewing environments into existing social media platforms  (amplifying the surveillance concerns outlined above). With this integration, AR environments might enable audiences to communicate back to the broadcaster and with each other, giving creators a better sense of how audiences react to and engage with their messages. Such feedback might also generate a limited degree of accountability, allowing creators to keep track of the impact and reach of their work, but also raising important questions around data privacy and control.

\subsubsection{Accountability and Co-option}With the recent hype and adjacent criticism surrounding the metaverse, an imagined virtual environment extending the capacities of AR, we find much overlooked around questions of accountability. Similar to prior metaverse instantiations like Second Life\cite{boellstorff2015coming}, the parallels drawn between the social media and AR, in terms of its trajectory \cite{barlow_2018} and potential \cite{zuckerman_2007}, raise questions about its propensity for co-option and expropriation \cite{nakamura2019virtual}. What would stop someone from using the same platforms designed for equity and justice initiatives as tools for hate speech? How does the lack of legal restrictions shape what corporate platforms might extract? These questions resonate with some notable elements we found in our data, such as P3's initiatives to challenge technocratic power, P15 reluctance to store Indigenous knowledge within certain platforms, and P18's bittersweet experience with collaborators. As creators continue to develop new and different ties to AR platforms, what allows those platforms to hold onto their liberatory potential lies less in the tool itself than the institutional conditions in which its situated. Creators' reflections suggest that what matters most is how those platforms are collectively managed and to whom they are accountable.

\subsubsection{Radical Change Driven by Minoritarian Groups}
Participants felt AR may be especially beneficial for activism projects focused on people disproportionally harmed along existing lines of inequity (race, class, gender, disability). By causing strong emotional reactions and overlaying key aspects of their own identities onto physical locations, creators believe that members of these groups could feel empowered and represented. Creators could center their experiences and feelings within the very environments that tend to shut them out. We note that a number of participants built experiences around marginalized groups, including recent AAPI struggles in New York (P10), 
repairing settler-Indigenous relationships in Canada (P15), creating monuments for historical Black figures in Los Angeles (P13), and creating formal and informal AR educational material about the historical contributions of people of color (P19).

We observed that the immersive potential of AR experiences encouraged a certain radical imagination among participants. Participants used AR tools to help people not only visualize what might be (making the invisible visible), but also realize the what might never be (making the impossible possible). The idea of making the cosmos the ground, and the ocean the sky, a notion we saw in both our pilot and main study data, came up repeatedly in interviews around the particular features of AR platforms that enable the revising of physical realities. ``I think it was very important for us to create this space where people can actively participate in that dreaming so it's like daydreaming. How we might extend our realities, our concepts of liberation \ldots outside of you know, our limits,'' a pilot study participant told us. 
To push against the limits of a physical reality mired in structural racism and oppression, participants found a subtle but profound sense of hope in the remaking of everyday landscapes. The dollar bill or the street sign stood for mundane pillars of a pervasive system of social control that could be meaningfully changed or subverted through artistic intervention.

This idea of revising the `impossible' undergirds some of the most compelling and celebrated interventions of 20th century, from the activism of Ida B. Wells to the artwork of James Luna. The fact that AR creators also play with this sensibility suggests an important aesthetic dimension to hybrid activism. By inviting people into alternative realities, AR authoring environments seemed to ignite creators’ radical imagination, drawing attention to the mutability of some of the most fundamental and seemingly permanent fixtures of our surrounding world.

\subsection{Limitations \& Future Work}
We interviewed a wide variety of participants with different types of AR activism projects, and learned how they view AR's potential as a tool to advocate for social causes. However, several aspects of our study design limit our findings. We outline these limitations and explore how they might be addressed in future work.

One limitation of our work is the difficulty of evaluating the impact and effectiveness of AR activism. Our assumptions and claims are based on creators' anecdotes and reports. Future work should consider interviewing or surveying audience members (i.e., people who viewed the creators' projects) and other stakeholders (e.g., property owners) to understand broader perceptions and experiences with AR activism. Additionally, participants were unaware of reliable ways to assess or measure their project's impact. This limitation is particularly problematic for researchers trying to understand the ability of AR activism to achieve social change. To address this shortfall, future work might run longer-term ethnographic fieldwork or controlled studies with both creators and audiences to test the potential for AR activism experiences to affect community-driven or prosocial outcomes, such as changed perspectives or attitudes, the strength and type of emotional reactions, or taking additional action.

A second limitation of this work involves the majority of non-disabled participants chosen for this study and the corresponding challenges around centering access. Even with the best intentions and investments in discoverability, AR faces several barriers to engagement and use that greatly limit its engagement with disability justice movements and intersectional advocacy \cite{herskovitz2020making,albouys2018towards}. Creators mentioned the hurdles to participation among groups without technical know-how, awareness, or familiarity. But they did not discuss the limitations associated with platforms that disproportionally exclude people with disabilities, as AR authors and audience members. The overwhelming emphasis of AR experiences discussed by participants was visual---relying on representations with few, if any, accommodations for blind users, a fact likely related to who and what we focused on in this study \cite{shinohara2018tenets,zhao2018enabling, herskovitz2020making}. These overlooked access issue suggest the need for additional diversity in the selection of interviewees, projects, and projects audiences to understand the limitations to radical imagination in its current visually-oriented form.

By interviewing a range of contemporary AR activists, our work has only begun to scratch the surface of their activity. As the technology and its adoption evolves, this work offers initial insights into creators' needs, and the potential opportunities and challenges of using AR for activism. Future work should build on this analysis by describing the broader design space of AR activism, including the impact of different contexts for activism, technologies or devices used, strategies for augmentation, audience perspectives on participation, and consequentially, considerations around platform ownership and maintenance in the creation of new AR tools for social change.

\section{Conclusion}
We contribute, to our knowledge, one of the first studies detailing creator's experiences with AR activism, including their decisions, expectations, and ability to achieve their goals. We interviewed twenty people who worked on projects that used AR for social change, ranging from augmenting objects to display truths about corruption, to augmenting public spaces to reveal hidden histories of marginalized groups. We found that AR can be a tool with low barrier to entry and potentially high impact in emotional reactions to messages. However, limitations in the technology and available platforms still restrict participants' abilities to create and distribute their created experiences to audiences. We discuss open questions and implications for researchers, activist communities, and AR platform developers to consider around AR activism.

\bibliographystyle{ACM-Reference-Format}
\bibliography{references}


\begin{thebibliography}{67}

%%% the Web address.

\ifx \showCODEN    \undefined \def \showCODEN     #1{\unskip}     \fi
\ifx \showDOI      \undefined \def \showDOI       #1{#1}\fi
\ifx \showISBNx    \undefined \def \showISBNx     #1{\unskip}     \fi
\ifx \showISBNxiii \undefined \def \showISBNxiii  #1{\unskip}     \fi
\ifx \showISSN     \undefined \def \showISSN      #1{\unskip}     \fi
\ifx \showLCCN     \undefined \def \showLCCN      #1{\unskip}     \fi
\ifx \shownote     \undefined \def \shownote      #1{#1}          \fi
\ifx \showarticletitle \undefined \def \showarticletitle #1{#1}   \fi
\ifx \showURL      \undefined \def \showURL       {\relax}        \fi
% invisible to TeX
\providecommand\bibfield[2]{#2}
\providecommand\bibinfo[2]{#2}
\providecommand\natexlab[1]{#1}
\providecommand\showeprint[2][]{arXiv:#2}

\bibitem[\protect\citeauthoryear{Albouys-Perrois, Laviole, Briant, and
  Brock}{Albouys-Perrois et~al\mbox{.}}{2018}]%
        {albouys2018towards}
\bibfield{author}{\bibinfo{person}{J{\'e}r{\'e}my Albouys-Perrois},
  \bibinfo{person}{J{\'e}r{\'e}my Laviole}, \bibinfo{person}{Carine Briant},
  {and} \bibinfo{person}{Anke~M Brock}.} \bibinfo{year}{2018}\natexlab{}.
\newblock \showarticletitle{Towards a multisensory augmented reality map for
  blind and low vision people: A participatory design approach}. In
  \bibinfo{booktitle}{\emph{Proceedings of the 2018 CHI conference on human
  factors in computing systems}}. \bibinfo{pages}{1--14}.
\newblock


\bibitem[\protect\citeauthoryear{Almond, McMahon, Janes, Whistance-Smith,
  Steinhauer, and Steinhauer}{Almond et~al\mbox{.}}{2018}]%
        {almond2018we}
\bibfield{author}{\bibinfo{person}{Amanda Almond}, \bibinfo{person}{Rob
  McMahon}, \bibinfo{person}{D Janes}, \bibinfo{person}{Greg Whistance-Smith},
  \bibinfo{person}{Diana Steinhauer}, {and} \bibinfo{person}{Stewart
  Steinhauer}.} \bibinfo{year}{2018}\natexlab{}.
\newblock \showarticletitle{We are all related: Using augmented reality as a
  learning resource for Indigenous-settler relations}.
\newblock \bibinfo{journal}{\emph{Northern Public Affairs}}
  \bibinfo{volume}{6}, \bibinfo{number}{2} (\bibinfo{year}{2018}).
\newblock


\bibitem[\protect\citeauthoryear{Aoyama and Aflatoony}{Aoyama and
  Aflatoony}{2020}]%
        {aoyama2020homemodar}
\bibfield{author}{\bibinfo{person}{Hiroo Aoyama} {and} \bibinfo{person}{Leila
  Aflatoony}.} \bibinfo{year}{2020}\natexlab{}.
\newblock \showarticletitle{HomeModAR: A Home Intervention Augmented Reality
  Tool for Occupational Therapists}. In \bibinfo{booktitle}{\emph{Extended
  Abstracts of the 2020 CHI Conference on Human Factors in Computing Systems}}.
  \bibinfo{pages}{1--7}.
\newblock


\bibitem[\protect\citeauthoryear{Barlow}{Barlow}{2018}]%
        {barlow_2018}
\bibfield{author}{\bibinfo{person}{John~Perry Barlow}.}
  \bibinfo{year}{2018}\natexlab{}.
\newblock \bibinfo{title}{A declaration of the independence of cyberspace}.
\newblock
\newblock
\urldef\tempurl%
\url{https://www.eff.org/cyberspace-independence}
\showURL{%
\tempurl}


\bibitem[\protect\citeauthoryear{Bennett and Segerberg}{Bennett and
  Segerberg}{2012}]%
        {bennett2012logic}
\bibfield{author}{\bibinfo{person}{W~Lance Bennett} {and}
  \bibinfo{person}{Alexandra Segerberg}.} \bibinfo{year}{2012}\natexlab{}.
\newblock \showarticletitle{The logic of connective action: Digital media and
  the personalization of contentious politics}.
\newblock \bibinfo{journal}{\emph{Information, communication \& society}}
  \bibinfo{volume}{15}, \bibinfo{number}{5} (\bibinfo{year}{2012}),
  \bibinfo{pages}{739--768}.
\newblock


\bibitem[\protect\citeauthoryear{Boellstorff}{Boellstorff}{2015}]%
        {boellstorff2015coming}
\bibfield{author}{\bibinfo{person}{Tom Boellstorff}.}
  \bibinfo{year}{2015}\natexlab{}.
\newblock \bibinfo{booktitle}{\emph{Coming of age in Second Life}}.
\newblock \bibinfo{publisher}{Princeton University Press}.
\newblock


\bibitem[\protect\citeauthoryear{Boyd and Mitchell}{Boyd and Mitchell}{2013}]%
        {boyd2013beautiful}
\bibfield{author}{\bibinfo{person}{Andrew Boyd} {and}
  \bibinfo{person}{David~Oswald Mitchell}.} \bibinfo{year}{2013}\natexlab{}.
\newblock \bibinfo{booktitle}{\emph{Beautiful Trouble: A Toolbox For Revolution
  (Pocket Edition)}}.
\newblock \bibinfo{publisher}{Or Books}.
\newblock


\bibitem[\protect\citeauthoryear{Costanza-Chock}{Costanza-Chock}{2020}]%
        {costanza2020design}
\bibfield{author}{\bibinfo{person}{Sasha Costanza-Chock}.}
  \bibinfo{year}{2020}\natexlab{}.
\newblock \bibinfo{booktitle}{\emph{Design justice: Community-led practices to
  build the worlds we need}}.
\newblock \bibinfo{publisher}{The MIT Press}.
\newblock


\bibitem[\protect\citeauthoryear{Criado}{Criado}{2020}]%
        {criado2020anthropology}
\bibfield{author}{\bibinfo{person}{Tom{\'a}s~S{\'a}nchez Criado}.}
  \bibinfo{year}{2020}\natexlab{}.
\newblock \showarticletitle{Anthropology as a careful design practice?}
\newblock \bibinfo{journal}{\emph{Zeitschrift fuer Ethnologie}}
  \bibinfo{volume}{145}, \bibinfo{number}{1} (\bibinfo{year}{2020}),
  \bibinfo{pages}{47--70}.
\newblock


\bibitem[\protect\citeauthoryear{CRL}{CRL}{[n.d.]}]%
        {creativereact}
\bibfield{author}{\bibinfo{person}{CRL}.} \bibinfo{year}{[n.d.]}\natexlab{}.
\newblock \bibinfo{title}{"Creative Reaction Lab"}.
\newblock \bibinfo{howpublished}{\url{https://www.creativereactionlab.com}}.
\newblock
\newblock
\shownote{Accessed: 2020-08-30.}


\bibitem[\protect\citeauthoryear{De~Choudhury, Jhaver, Sugar, and
  Weber}{De~Choudhury et~al\mbox{.}}{2016}]%
        {de2016social}
\bibfield{author}{\bibinfo{person}{Munmun De~Choudhury},
  \bibinfo{person}{Shagun Jhaver}, \bibinfo{person}{Benjamin Sugar}, {and}
  \bibinfo{person}{Ingmar Weber}.} \bibinfo{year}{2016}\natexlab{}.
\newblock \showarticletitle{Social media participation in an activist movement
  for racial equality}. In \bibinfo{booktitle}{\emph{Tenth International AAAI
  Conference on Web and Social Media}}.
\newblock


\bibitem[\protect\citeauthoryear{De~S{\'a} and Churchill}{De~S{\'a} and
  Churchill}{2013}]%
        {de2013mobile}
\bibfield{author}{\bibinfo{person}{Marco De~S{\'a}} {and}
  \bibinfo{person}{Elizabeth~F Churchill}.} \bibinfo{year}{2013}\natexlab{}.
\newblock \showarticletitle{Mobile augmented reality: A design perspective}.
\newblock In \bibinfo{booktitle}{\emph{Human factors in augmented reality
  environments}}. \bibinfo{publisher}{Springer}, \bibinfo{pages}{139--164}.
\newblock


\bibitem[\protect\citeauthoryear{Dimond, Dye, LaRose, and Bruckman}{Dimond
  et~al\mbox{.}}{2013}]%
        {dimond2013hollaback}
\bibfield{author}{\bibinfo{person}{Jill~P Dimond}, \bibinfo{person}{Michaelanne
  Dye}, \bibinfo{person}{Daphne LaRose}, {and} \bibinfo{person}{Amy~S
  Bruckman}.} \bibinfo{year}{2013}\natexlab{}.
\newblock \showarticletitle{Hollaback! The role of storytelling online in a
  social movement organization}. In \bibinfo{booktitle}{\emph{Proceedings of
  the 2013 conference on Computer supported cooperative work}}.
  \bibinfo{pages}{477--490}.
\newblock


\bibitem[\protect\citeauthoryear{DiSalvo, Lukens, Lodato, Jenkins, and
  Kim}{DiSalvo et~al\mbox{.}}{2014}]%
        {disalvo2014making}
\bibfield{author}{\bibinfo{person}{Carl DiSalvo}, \bibinfo{person}{Jonathan
  Lukens}, \bibinfo{person}{Thomas Lodato}, \bibinfo{person}{Tom Jenkins},
  {and} \bibinfo{person}{Tanyoung Kim}.} \bibinfo{year}{2014}\natexlab{}.
\newblock \showarticletitle{Making public things: how HCI design can express
  matters of concern}. In \bibinfo{booktitle}{\emph{Proceedings of the SIGCHI
  Conference on Human Factors in Computing Systems}}.
  \bibinfo{pages}{2397--2406}.
\newblock


\bibitem[\protect\citeauthoryear{Farber, Crombie, Ortiz, Shabazz, and
  Williamson}{Farber et~al\mbox{.}}{2019}]%
        {farber2019monument}
\bibfield{author}{\bibinfo{person}{Paul Farber}, \bibinfo{person}{Kara
  Crombie}, \bibinfo{person}{Michelle~Angela Ortiz}, \bibinfo{person}{Jamel
  Shabazz}, {and} \bibinfo{person}{Marisa Williamson}.}
  \bibinfo{year}{2019}\natexlab{}.
\newblock \showarticletitle{" Monument Lab: Prototypes/Proposals" Installation
  Image}.
\newblock  (\bibinfo{year}{2019}).
\newblock


\bibitem[\protect\citeauthoryear{Fisher}{Fisher}{2021}]%
        {fisher2021augmented}
\bibfield{author}{\bibinfo{person}{Joshua~A Fisher}.}
  \bibinfo{year}{2021}\natexlab{}.
\newblock \bibinfo{booktitle}{\emph{Augmented and Mixed Reality for
  Communities}}.
\newblock \bibinfo{publisher}{CRC Press}.
\newblock


\bibitem[\protect\citeauthoryear{Fox, Lim, Hirsch, and Rosner}{Fox
  et~al\mbox{.}}{2020}]%
        {fox2020accounting}
\bibfield{author}{\bibinfo{person}{Sarah Fox}, \bibinfo{person}{Catherine Lim},
  \bibinfo{person}{Tad Hirsch}, {and} \bibinfo{person}{Daniela~K Rosner}.}
  \bibinfo{year}{2020}\natexlab{}.
\newblock \showarticletitle{Accounting for Design Activism: On the
  Positionality and Politics of Designerly Intervention}.
\newblock \bibinfo{journal}{\emph{Design Issues}} \bibinfo{volume}{36},
  \bibinfo{number}{1} (\bibinfo{year}{2020}), \bibinfo{pages}{5--18}.
\newblock


\bibitem[\protect\citeauthoryear{Fox, Silva, and Rosner}{Fox
  et~al\mbox{.}}{2018}]%
        {fox2018beyond}
\bibfield{author}{\bibinfo{person}{Sarah~E Fox}, \bibinfo{person}{Rafael~ML
  Silva}, {and} \bibinfo{person}{Daniela~K Rosner}.}
  \bibinfo{year}{2018}\natexlab{}.
\newblock \showarticletitle{Beyond the prototype: Maintenance, collective
  responsibility, and public IoT}. In \bibinfo{booktitle}{\emph{Proceedings of
  the 2018 Designing Interactive Systems Conference}}. \bibinfo{pages}{21--32}.
\newblock


\bibitem[\protect\citeauthoryear{Freeman}{Freeman}{2012}]%
        {freeman2012manifestar}
\bibfield{author}{\bibinfo{person}{John~Craig Freeman}.}
  \bibinfo{year}{2012}\natexlab{}.
\newblock \showarticletitle{ManifestAR: an augmented reality manifesto}. In
  \bibinfo{booktitle}{\emph{The Engineering Reality of Virtual Reality 2012}},
  Vol.~\bibinfo{volume}{8289}. International Society for Optics and Photonics,
  \bibinfo{pages}{82890D}.
\newblock


\bibitem[\protect\citeauthoryear{Fuad-Luke}{Fuad-Luke}{2013}]%
        {fuad2013design}
\bibfield{author}{\bibinfo{person}{Alastair Fuad-Luke}.}
  \bibinfo{year}{2013}\natexlab{}.
\newblock \bibinfo{booktitle}{\emph{Design activism: beautiful strangeness for
  a sustainable world}}.
\newblock \bibinfo{publisher}{Routledge}.
\newblock


\bibitem[\protect\citeauthoryear{George and Leidner}{George and
  Leidner}{2019}]%
        {george2019clicktivism}
\bibfield{author}{\bibinfo{person}{Jordana~J George} {and}
  \bibinfo{person}{Dorothy~E Leidner}.} \bibinfo{year}{2019}\natexlab{}.
\newblock \showarticletitle{From clicktivism to hacktivism: Understanding
  digital activism}.
\newblock \bibinfo{journal}{\emph{Information and Organization}}
  \bibinfo{volume}{29}, \bibinfo{number}{3} (\bibinfo{year}{2019}),
  \bibinfo{pages}{100249}.
\newblock


\bibitem[\protect\citeauthoryear{Gleason}{Gleason}{2013}]%
        {gleason2013occupy}
\bibfield{author}{\bibinfo{person}{Benjamin Gleason}.}
  \bibinfo{year}{2013}\natexlab{}.
\newblock \showarticletitle{\# Occupy Wall Street: Exploring informal learning
  about a social movement on Twitter}.
\newblock \bibinfo{journal}{\emph{American Behavioral Scientist}}
  \bibinfo{volume}{57}, \bibinfo{number}{7} (\bibinfo{year}{2013}),
  \bibinfo{pages}{966--982}.
\newblock


\bibitem[\protect\citeauthoryear{Gonsalves, Foth, Caldwell, and
  Jenek}{Gonsalves et~al\mbox{.}}{2021}]%
        {gonsalves2021radical}
\bibfield{author}{\bibinfo{person}{Kavita Gonsalves}, \bibinfo{person}{Marcus
  Foth}, \bibinfo{person}{Glenda Caldwell}, {and} \bibinfo{person}{Waldemar
  Jenek}.} \bibinfo{year}{2021}\natexlab{}.
\newblock \showarticletitle{Radical placemaking: An immersive, experiential and
  activist approach for marginalised communities}.
\newblock \bibinfo{journal}{\emph{Connections: Exploring heritage,
  architecture, cities, art, media. Vol. 20.1.}} (\bibinfo{year}{2021}),
  \bibinfo{pages}{237--252}.
\newblock


\bibitem[\protect\citeauthoryear{Guo, Canberk, Murphy, Monroy-Hern{\'a}ndez,
  and Vaish}{Guo et~al\mbox{.}}{2019}]%
        {guo2019blocks}
\bibfield{author}{\bibinfo{person}{Anhong Guo}, \bibinfo{person}{Ilter
  Canberk}, \bibinfo{person}{Hannah Murphy}, \bibinfo{person}{Andr{\'e}s
  Monroy-Hern{\'a}ndez}, {and} \bibinfo{person}{Rajan Vaish}.}
  \bibinfo{year}{2019}\natexlab{}.
\newblock \showarticletitle{Blocks: Collaborative and persistent augmented
  reality experiences}.
\newblock \bibinfo{journal}{\emph{Proceedings of the ACM on Interactive,
  Mobile, Wearable and Ubiquitous Technologies}} \bibinfo{volume}{3},
  \bibinfo{number}{3} (\bibinfo{year}{2019}), \bibinfo{pages}{1--24}.
\newblock


\bibitem[\protect\citeauthoryear{Harley}{Harley}{2020}]%
        {harley_2020}
\bibfield{author}{\bibinfo{person}{Aurora Harley}.}
  \bibinfo{year}{2020}\natexlab{}.
\newblock \bibinfo{title}{UX guidelines for augmented-reality shopping tools}.
\newblock
\newblock
\urldef\tempurl%
\url{https://www.nngroup.com/articles/augmented-reality-ecommerce-guidelines/}
\showURL{%
\tempurl}


\bibitem[\protect\citeauthoryear{Heaney and Rojas}{Heaney and Rojas}{2014}]%
        {heaney2014hybrid}
\bibfield{author}{\bibinfo{person}{Michael~T Heaney} {and}
  \bibinfo{person}{Fabio Rojas}.} \bibinfo{year}{2014}\natexlab{}.
\newblock \showarticletitle{Hybrid activism: Social movement mobilization in a
  multimovement environment}.
\newblock \bibinfo{journal}{\emph{Amer. J. Sociology}} \bibinfo{volume}{119},
  \bibinfo{number}{4} (\bibinfo{year}{2014}), \bibinfo{pages}{1047--1103}.
\newblock


\bibitem[\protect\citeauthoryear{Herrera, Bailenson, Weisz, Ogle, and
  Zaki}{Herrera et~al\mbox{.}}{2018}]%
        {herrera2018building}
\bibfield{author}{\bibinfo{person}{Fernanda Herrera}, \bibinfo{person}{Jeremy
  Bailenson}, \bibinfo{person}{Erika Weisz}, \bibinfo{person}{Elise Ogle},
  {and} \bibinfo{person}{Jamil Zaki}.} \bibinfo{year}{2018}\natexlab{}.
\newblock \showarticletitle{Building long-term empathy: A large-scale
  comparison of traditional and virtual reality perspective-taking}.
\newblock \bibinfo{journal}{\emph{PloS one}} \bibinfo{volume}{13},
  \bibinfo{number}{10} (\bibinfo{year}{2018}), \bibinfo{pages}{e0204494}.
\newblock


\bibitem[\protect\citeauthoryear{Herskovitz, Wu, White, Pavel, Reyes, Guo, and
  Bigham}{Herskovitz et~al\mbox{.}}{2020}]%
        {herskovitz2020making}
\bibfield{author}{\bibinfo{person}{Jaylin Herskovitz}, \bibinfo{person}{Jason
  Wu}, \bibinfo{person}{Samuel White}, \bibinfo{person}{Amy Pavel},
  \bibinfo{person}{Gabriel Reyes}, \bibinfo{person}{Anhong Guo}, {and}
  \bibinfo{person}{Jeffrey~P Bigham}.} \bibinfo{year}{2020}\natexlab{}.
\newblock \showarticletitle{Making mobile augmented reality applications
  accessible}. In \bibinfo{booktitle}{\emph{The 22nd International ACM
  SIGACCESS Conference on Computers and Accessibility}}.
  \bibinfo{pages}{1--14}.
\newblock


\bibitem[\protect\citeauthoryear{Hirsch and Henry}{Hirsch and Henry}{2005}]%
        {hirsch2005txtmob}
\bibfield{author}{\bibinfo{person}{Tad Hirsch} {and} \bibinfo{person}{John
  Henry}.} \bibinfo{year}{2005}\natexlab{}.
\newblock \showarticletitle{TXTmob: text messaging for protest swarms}. In
  \bibinfo{booktitle}{\emph{CHI'05 extended abstracts on Human factors in
  computing systems}}. \bibinfo{pages}{1455--1458}.
\newblock


\bibitem[\protect\citeauthoryear{Jurgenson}{Jurgenson}{2011}]%
        {jurgenson_2011}
\bibfield{author}{\bibinfo{person}{Nathan Jurgenson}.}
  \bibinfo{year}{2011}\natexlab{}.
\newblock \bibinfo{title}{Digital Dualism versus Augmented Reality}.
\newblock
\newblock
\urldef\tempurl%
\url{https://thesocietypages.org/cyborgology/2011/02/24/digital-dualism-versus-augmented-reality/}
\showURL{%
\tempurl}


\bibitem[\protect\citeauthoryear{Jylh{\"a}, Nurmi, Sir{\'e}n, Hemminki, and
  Jacucci}{Jylh{\"a} et~al\mbox{.}}{2013}]%
        {jylha2013matkahupi}
\bibfield{author}{\bibinfo{person}{Antti Jylh{\"a}}, \bibinfo{person}{Petteri
  Nurmi}, \bibinfo{person}{Miika Sir{\'e}n}, \bibinfo{person}{Samuli Hemminki},
  {and} \bibinfo{person}{Giulio Jacucci}.} \bibinfo{year}{2013}\natexlab{}.
\newblock \showarticletitle{Matkahupi: a persuasive mobile application for
  sustainable mobility}. In \bibinfo{booktitle}{\emph{Proceedings of the 2013
  ACM conference on Pervasive and ubiquitous computing adjunct publication}}.
  \bibinfo{pages}{227--230}.
\newblock


\bibitem[\protect\citeauthoryear{Kors, Ferri, Van Der~Spek, Ketel, and
  Schouten}{Kors et~al\mbox{.}}{2016}]%
        {kors2016breathtaking}
\bibfield{author}{\bibinfo{person}{Martijn~JL Kors}, \bibinfo{person}{Gabriele
  Ferri}, \bibinfo{person}{Erik~D Van Der~Spek}, \bibinfo{person}{Cas Ketel},
  {and} \bibinfo{person}{Ben~AM Schouten}.} \bibinfo{year}{2016}\natexlab{}.
\newblock \showarticletitle{A breathtaking journey. On the design of an
  empathy-arousing mixed-reality game}. In
  \bibinfo{booktitle}{\emph{Proceedings of the 2016 Annual Symposium on
  Computer-Human Interaction in Play}}. \bibinfo{pages}{91--104}.
\newblock


\bibitem[\protect\citeauthoryear{Kuznetsov, Davis, Paulos, Gross, and
  Cheung}{Kuznetsov et~al\mbox{.}}{2011}]%
        {kuznetsov2011red}
\bibfield{author}{\bibinfo{person}{Stacey Kuznetsov},
  \bibinfo{person}{George~Noel Davis}, \bibinfo{person}{Eric Paulos},
  \bibinfo{person}{Mark~D Gross}, {and} \bibinfo{person}{Jian~Chiu Cheung}.}
  \bibinfo{year}{2011}\natexlab{}.
\newblock \showarticletitle{Red balloon, green balloon, sensors in the sky}. In
  \bibinfo{booktitle}{\emph{Proceedings of the 13th international conference on
  Ubiquitous computing}}. \bibinfo{pages}{237--246}.
\newblock


\bibitem[\protect\citeauthoryear{Kuznetsov, Paulos, and Gross}{Kuznetsov
  et~al\mbox{.}}{2010}]%
        {kuznetsov2010wallbots}
\bibfield{author}{\bibinfo{person}{Stacey Kuznetsov}, \bibinfo{person}{Eric
  Paulos}, {and} \bibinfo{person}{Mark~D Gross}.}
  \bibinfo{year}{2010}\natexlab{}.
\newblock \showarticletitle{WallBots: interactive wall-crawling robots in the
  hands of public artists and political activists}. In
  \bibinfo{booktitle}{\emph{Proceedings of the 8th ACM Conference on Designing
  Interactive Systems}}. \bibinfo{pages}{208--217}.
\newblock


\bibitem[\protect\citeauthoryear{Le~Dantec}{Le~Dantec}{2016}]%
        {le2016designing}
\bibfield{author}{\bibinfo{person}{Christopher~A Le~Dantec}.}
  \bibinfo{year}{2016}\natexlab{}.
\newblock \bibinfo{booktitle}{\emph{Designing publics}}.
\newblock \bibinfo{publisher}{MIT Press}.
\newblock


\bibitem[\protect\citeauthoryear{Lee and Hsieh}{Lee and Hsieh}{2013}]%
        {lee2013does}
\bibfield{author}{\bibinfo{person}{Yu-Hao Lee} {and} \bibinfo{person}{Gary
  Hsieh}.} \bibinfo{year}{2013}\natexlab{}.
\newblock \showarticletitle{Does slacktivism hurt activism? The effects of
  moral balancing and consistency in online activism}. In
  \bibinfo{booktitle}{\emph{Proceedings of the SIGCHI conference on human
  factors in computing systems}}. \bibinfo{pages}{811--820}.
\newblock


\bibitem[\protect\citeauthoryear{Li, Bora, Salvi, and Brady}{Li
  et~al\mbox{.}}{2018}]%
        {li2018slacktivists}
\bibfield{author}{\bibinfo{person}{Hanlin Li}, \bibinfo{person}{Disha Bora},
  \bibinfo{person}{Sagar Salvi}, {and} \bibinfo{person}{Erin Brady}.}
  \bibinfo{year}{2018}\natexlab{}.
\newblock \showarticletitle{Slacktivists or Activists? Identity Work in the
  Virtual Disability March}. In \bibinfo{booktitle}{\emph{Proceedings of the
  2018 CHI Conference on Human Factors in Computing Systems}}.
  \bibinfo{pages}{1--13}.
\newblock


\bibitem[\protect\citeauthoryear{Liu, Ford, Parnin, and Dabbish}{Liu
  et~al\mbox{.}}{2017}]%
        {liu2017selfies}
\bibfield{author}{\bibinfo{person}{Fannie Liu}, \bibinfo{person}{Denae Ford},
  \bibinfo{person}{Chris Parnin}, {and} \bibinfo{person}{Laura Dabbish}.}
  \bibinfo{year}{2017}\natexlab{}.
\newblock \showarticletitle{Selfies as social movements: Influences on
  participation and perceived impact on stereotypes}.
\newblock \bibinfo{journal}{\emph{Proceedings of the ACM on Human-Computer
  Interaction}} \bibinfo{volume}{1}, \bibinfo{number}{CSCW}
  (\bibinfo{year}{2017}), \bibinfo{pages}{1--21}.
\newblock


\bibitem[\protect\citeauthoryear{Maclntyre and Smith}{Maclntyre and
  Smith}{2018}]%
        {maclntyre2018thoughts}
\bibfield{author}{\bibinfo{person}{Blair Maclntyre} {and}
  \bibinfo{person}{Trevor~F Smith}.} \bibinfo{year}{2018}\natexlab{}.
\newblock \showarticletitle{Thoughts on the Future of WebXR and the Immersive
  Web}. In \bibinfo{booktitle}{\emph{2018 IEEE International Symposium on Mixed
  and Augmented Reality Adjunct (ISMAR-Adjunct)}}. IEEE,
  \bibinfo{pages}{338--342}.
\newblock


\bibitem[\protect\citeauthoryear{Markussen}{Markussen}{2013}]%
        {markussen2013disruptive}
\bibfield{author}{\bibinfo{person}{Thomas Markussen}.}
  \bibinfo{year}{2013}\natexlab{}.
\newblock \showarticletitle{The disruptive aesthetics of design activism:
  enacting design between art and politics}.
\newblock \bibinfo{journal}{\emph{Design Issues}} \bibinfo{volume}{29},
  \bibinfo{number}{1} (\bibinfo{year}{2013}), \bibinfo{pages}{38--50}.
\newblock


\bibitem[\protect\citeauthoryear{McMahon, Almond, Whistance-Smith, Steinhauer,
  Steinhauer, and Janes}{McMahon et~al\mbox{.}}{2019}]%
        {mcmahon2019sweetgrass}
\bibfield{author}{\bibinfo{person}{Rob McMahon}, \bibinfo{person}{Amanda
  Almond}, \bibinfo{person}{Greg Whistance-Smith}, \bibinfo{person}{Diana
  Steinhauer}, \bibinfo{person}{Stewart Steinhauer}, {and}
  \bibinfo{person}{Diane~P Janes}.} \bibinfo{year}{2019}\natexlab{}.
\newblock \showarticletitle{Sweetgrass AR: Exploring augmented reality as a
  resource for Indigenous--settler relations}.
\newblock \bibinfo{journal}{\emph{International Journal of Communication}}
  \bibinfo{volume}{13} (\bibinfo{year}{2019}), \bibinfo{pages}{23}.
\newblock


\bibitem[\protect\citeauthoryear{Milo{\v{s}}evi{\'c}-{\DJ}or{\dj}evi{\'c} and
  {\v{Z}}e{\v{z}}elj}{Milo{\v{s}}evi{\'c}-{\DJ}or{\dj}evi{\'c} and
  {\v{Z}}e{\v{z}}elj}{2017}]%
        {milovsevic2017civic}
\bibfield{author}{\bibinfo{person}{Jasna~S
  Milo{\v{s}}evi{\'c}-{\DJ}or{\dj}evi{\'c}} {and} \bibinfo{person}{Iris~L
  {\v{Z}}e{\v{z}}elj}.} \bibinfo{year}{2017}\natexlab{}.
\newblock \showarticletitle{Civic activism online: Making young people dormant
  or more active in real life?}
\newblock \bibinfo{journal}{\emph{Computers in Human Behavior}}
  \bibinfo{volume}{70} (\bibinfo{year}{2017}), \bibinfo{pages}{113--118}.
\newblock


\bibitem[\protect\citeauthoryear{Monroy-Hern{\'a}ndez, Boyd, Kiciman,
  De~Choudhury, and Counts}{Monroy-Hern{\'a}ndez et~al\mbox{.}}{2013}]%
        {monroy2013new}
\bibfield{author}{\bibinfo{person}{Andr{\'e}s Monroy-Hern{\'a}ndez},
  \bibinfo{person}{Danah Boyd}, \bibinfo{person}{Emre Kiciman},
  \bibinfo{person}{Munmun De~Choudhury}, {and} \bibinfo{person}{Scott Counts}.}
  \bibinfo{year}{2013}\natexlab{}.
\newblock \showarticletitle{The new war correspondents: The rise of civic media
  curation in urban warfare}. In \bibinfo{booktitle}{\emph{Proceedings of the
  2013 conference on Computer supported cooperative work}}.
  \bibinfo{pages}{1443--1452}.
\newblock


\bibitem[\protect\citeauthoryear{Morozov}{Morozov}{2009}]%
        {morozov:bravenewworld}
\bibfield{author}{\bibinfo{person}{Evgeny Morozov}.}
  \bibinfo{year}{2009}\natexlab{}.
\newblock \bibinfo{title}{The brave new world of slacktivism}.
\newblock \bibinfo{howpublished}{\textit{Foreign Policy}}.
\newblock
\newblock
\shownote{Retrieved July 28, 2016 from
  http://www.npr.org/templates/story/story.php?storyId=104302141.}


\bibitem[\protect\citeauthoryear{Mostajeran, Steinicke, Ariza~Nunez, Gatsios,
  and Fotiadis}{Mostajeran et~al\mbox{.}}{2020}]%
        {mostajeran2020augmented}
\bibfield{author}{\bibinfo{person}{Fariba Mostajeran}, \bibinfo{person}{Frank
  Steinicke}, \bibinfo{person}{Oscar~Javier Ariza~Nunez},
  \bibinfo{person}{Dimitrios Gatsios}, {and} \bibinfo{person}{Dimitrios
  Fotiadis}.} \bibinfo{year}{2020}\natexlab{}.
\newblock \showarticletitle{Augmented reality for older adults: Exploring
  acceptability of virtual coaches for home-based balance training in an aging
  population}. In \bibinfo{booktitle}{\emph{Proceedings of the 2020 CHI
  Conference on Human Factors in Computing Systems}}. \bibinfo{pages}{1--12}.
\newblock


\bibitem[\protect\citeauthoryear{Nakamura}{Nakamura}{2019}]%
        {nakamura2019virtual}
\bibfield{author}{\bibinfo{person}{Lisa Nakamura}.}
  \bibinfo{year}{2019}\natexlab{}.
\newblock \showarticletitle{Virtual reality and the feeling of virtue: Women of
  color narrators, enforced hospitality, and the leveraging of empathy}. In
  \bibinfo{booktitle}{\emph{Proceedings of the 2019 on Designing Interactive
  Systems Conference}}. \bibinfo{pages}{3--3}.
\newblock


\bibitem[\protect\citeauthoryear{Paavilainen, Korhonen, Alha, Stenros,
  Koskinen, and Mayra}{Paavilainen et~al\mbox{.}}{2017}]%
        {paavilainen2017pokemon}
\bibfield{author}{\bibinfo{person}{Janne Paavilainen}, \bibinfo{person}{Hannu
  Korhonen}, \bibinfo{person}{Kati Alha}, \bibinfo{person}{Jaakko Stenros},
  \bibinfo{person}{Elina Koskinen}, {and} \bibinfo{person}{Frans Mayra}.}
  \bibinfo{year}{2017}\natexlab{}.
\newblock \showarticletitle{The Pok{\'e}mon GO experience: A location-based
  augmented reality mobile game goes mainstream}. In
  \bibinfo{booktitle}{\emph{Proceedings of the 2017 CHI conference on human
  factors in computing systems}}. \bibinfo{pages}{2493--2498}.
\newblock


\bibitem[\protect\citeauthoryear{Paulos, Honicky, and Hooker}{Paulos
  et~al\mbox{.}}{2009}]%
        {paulos2009citizen}
\bibfield{author}{\bibinfo{person}{Eric Paulos}, \bibinfo{person}{RJ Honicky},
  {and} \bibinfo{person}{Ben Hooker}.} \bibinfo{year}{2009}\natexlab{}.
\newblock \showarticletitle{Citizen science: Enabling participatory urbanism}.
\newblock In \bibinfo{booktitle}{\emph{Handbook of research on urban
  informatics: The practice and promise of the real-time city}}.
  \bibinfo{publisher}{IGI Global}, \bibinfo{pages}{414--436}.
\newblock


\bibitem[\protect\citeauthoryear{Piumsomboon, Lee, Irlitti, Ens, Thomas, and
  Billinghurst}{Piumsomboon et~al\mbox{.}}{2019}]%
        {piumsomboon2019shoulder}
\bibfield{author}{\bibinfo{person}{Thammathip Piumsomboon},
  \bibinfo{person}{Gun~A Lee}, \bibinfo{person}{Andrew Irlitti},
  \bibinfo{person}{Barrett Ens}, \bibinfo{person}{Bruce~H Thomas}, {and}
  \bibinfo{person}{Mark Billinghurst}.} \bibinfo{year}{2019}\natexlab{}.
\newblock \showarticletitle{On the shoulder of the giant: A multi-scale mixed
  reality collaboration with 360 video sharing and tangible interaction}. In
  \bibinfo{booktitle}{\emph{Proceedings of the 2019 CHI conference on human
  factors in computing systems}}. \bibinfo{pages}{1--17}.
\newblock


\bibitem[\protect\citeauthoryear{Project}{Project}{2017}]%
        {WholeStoryProject}
\bibfield{author}{\bibinfo{person}{The Whole~Story Project}.}
  \bibinfo{year}{2017}\natexlab{}.
\newblock \bibinfo{title}{WholeStoryProject}.
\newblock
\newblock
\urldef\tempurl%
\url{https://www.commarts.com/project/11275/the-whole-story-project-ar-app}
\showURL{%
Retrieved January 1, 2022 from \tempurl}


\bibitem[\protect\citeauthoryear{Refratik}{Refratik}{2015}]%
        {HackDaPatria}
\bibfield{author}{\bibinfo{person}{Refratik}.} \bibinfo{year}{2015}\natexlab{}.
\newblock \bibinfo{title}{HackDaPatriaProject}.
\newblock
\newblock
\urldef\tempurl%
\url{https://refrakt.org/states-of-mind}
\showURL{%
Retrieved January 1, 2022 from \tempurl}


\bibitem[\protect\citeauthoryear{Salim and Haque}{Salim and Haque}{2015}]%
        {salim2015urban}
\bibfield{author}{\bibinfo{person}{Flora Salim} {and} \bibinfo{person}{Usman
  Haque}.} \bibinfo{year}{2015}\natexlab{}.
\newblock \showarticletitle{Urban computing in the wild: A survey on large
  scale participation and citizen engagement with ubiquitous computing, cyber
  physical systems, and Internet of Things}.
\newblock \bibinfo{journal}{\emph{International Journal of Human-Computer
  Studies}}  \bibinfo{volume}{81} (\bibinfo{year}{2015}),
  \bibinfo{pages}{31--48}.
\newblock


\bibitem[\protect\citeauthoryear{Samuels, Mathew, Kommanivanh, Kwon, Gomez,
  Thunder, Velazquez, Martinez, and LaQueens}{Samuels et~al\mbox{.}}{2018}]%
        {10.1145/3170427.3186526}
\bibfield{author}{\bibinfo{person}{Janice~Tisha Samuels},
  \bibinfo{person}{Anijo~P. Mathew}, \bibinfo{person}{Chantala Kommanivanh},
  \bibinfo{person}{Daniel Kwon}, \bibinfo{person}{Liz Gomez},
  \bibinfo{person}{B'Rael~Ali Thunder}, \bibinfo{person}{Daria Velazquez},
  \bibinfo{person}{Millie Martinez}, {and} \bibinfo{person}{Leah LaQueens}.}
  \bibinfo{year}{2018}\natexlab{}.
\newblock \showarticletitle{Art, Human Computer Interaction, and Shared
  Experiences: A Gun Violence Prevention Intervention}. In
  \bibinfo{booktitle}{\emph{Extended Abstracts of the 2018 CHI Conference on
  Human Factors in Computing Systems}} (Montreal QC, Canada)
  \emph{(\bibinfo{series}{CHI EA '18})}. \bibinfo{publisher}{Association for
  Computing Machinery}, \bibinfo{address}{New York, NY, USA},
  \bibinfo{pages}{1–4}.
\newblock
\showISBNx{9781450356213}
\urldef\tempurl%
\url{https://doi.org/10.1145/3170427.3186526}
\showDOI{\tempurl}


\bibitem[\protect\citeauthoryear{Sanaeipoor and Emami}{Sanaeipoor and
  Emami}{2020}]%
        {sanaeipoor2020smart}
\bibfield{author}{\bibinfo{person}{Samaneh Sanaeipoor} {and}
  \bibinfo{person}{Khashayar~Hojjati Emami}.} \bibinfo{year}{2020}\natexlab{}.
\newblock \showarticletitle{Smart City: Exploring the Role of Augmented Reality
  in Placemaking}. In \bibinfo{booktitle}{\emph{2020 4th International
  Conference on Smart City, Internet of Things and Applications (SCIOT)}}.
  IEEE, \bibinfo{pages}{91--98}.
\newblock


\bibitem[\protect\citeauthoryear{Schmalstieg, Langlotz, and
  Billinghurst}{Schmalstieg et~al\mbox{.}}{2008}]%
        {Schmalstieg2008AugmentedR2}
\bibfield{author}{\bibinfo{person}{Dieter Schmalstieg}, \bibinfo{person}{Tobias
  Langlotz}, {and} \bibinfo{person}{Mark Billinghurst}.}
  \bibinfo{year}{2008}\natexlab{}.
\newblock \showarticletitle{Augmented Reality 2.0}. In
  \bibinfo{booktitle}{\emph{Virtual Realities}}.
\newblock


\bibitem[\protect\citeauthoryear{Shinohara, Bennett, Pratt, and
  Wobbrock}{Shinohara et~al\mbox{.}}{2018}]%
        {shinohara2018tenets}
\bibfield{author}{\bibinfo{person}{Kristen Shinohara},
  \bibinfo{person}{Cynthia~L Bennett}, \bibinfo{person}{Wanda Pratt}, {and}
  \bibinfo{person}{Jacob~O Wobbrock}.} \bibinfo{year}{2018}\natexlab{}.
\newblock \showarticletitle{Tenets for social accessibility: Towards humanizing
  disabled people in design}.
\newblock \bibinfo{journal}{\emph{ACM Transactions on Accessible Computing
  (TACCESS)}} \bibinfo{volume}{11}, \bibinfo{number}{1} (\bibinfo{year}{2018}),
  \bibinfo{pages}{1--31}.
\newblock


\bibitem[\protect\citeauthoryear{Skwarek}{Skwarek}{2018}]%
        {skwarek2018augmented}
\bibfield{author}{\bibinfo{person}{Mark Skwarek}.}
  \bibinfo{year}{2018}\natexlab{}.
\newblock \showarticletitle{Augmented reality activism}.
\newblock In \bibinfo{booktitle}{\emph{Augmented reality art}}.
  \bibinfo{publisher}{Springer}, \bibinfo{pages}{3--40}.
\newblock


\bibitem[\protect\citeauthoryear{Sutherland}{Sutherland}{1968}]%
        {sutherland1968head}
\bibfield{author}{\bibinfo{person}{Ivan~E Sutherland}.}
  \bibinfo{year}{1968}\natexlab{}.
\newblock \showarticletitle{A head-mounted three dimensional display}. In
  \bibinfo{booktitle}{\emph{Proceedings of the December 9-11, 1968, fall joint
  computer conference, part I}}. \bibinfo{pages}{757--764}.
\newblock


\bibitem[\protect\citeauthoryear{Thomas}{Thomas}{2006}]%
        {thomas2006general}
\bibfield{author}{\bibinfo{person}{David~R Thomas}.}
  \bibinfo{year}{2006}\natexlab{}.
\newblock \showarticletitle{A general inductive approach for analyzing
  qualitative evaluation data}.
\newblock \bibinfo{journal}{\emph{American journal of evaluation}}
  \bibinfo{volume}{27}, \bibinfo{number}{2} (\bibinfo{year}{2006}),
  \bibinfo{pages}{237--246}.
\newblock


\bibitem[\protect\citeauthoryear{Thorpe}{Thorpe}{2012}]%
        {thorpe2012architecture}
\bibfield{author}{\bibinfo{person}{Ann Thorpe}.}
  \bibinfo{year}{2012}\natexlab{}.
\newblock \bibinfo{booktitle}{\emph{Architecture and design versus consumerism:
  how design activism confronts growth}}.
\newblock \bibinfo{publisher}{Routledge}.
\newblock


\bibitem[\protect\citeauthoryear{Tran~O'Leary, Zewde, Mankoff, and
  Rosner}{Tran~O'Leary et~al\mbox{.}}{2019}]%
        {tran2019gets}
\bibfield{author}{\bibinfo{person}{Jasper Tran~O'Leary}, \bibinfo{person}{Sara
  Zewde}, \bibinfo{person}{Jennifer Mankoff}, {and} \bibinfo{person}{Daniela~K
  Rosner}.} \bibinfo{year}{2019}\natexlab{}.
\newblock \showarticletitle{Who gets to future? Race, representation, and
  design methods in Africatown}. In \bibinfo{booktitle}{\emph{Proceedings of
  the 2019 CHI Conference on Human Factors in Computing Systems}}.
  \bibinfo{pages}{1--13}.
\newblock


\bibitem[\protect\citeauthoryear{Tufekci and Wilson}{Tufekci and
  Wilson}{2012}]%
        {tufekci2012social}
\bibfield{author}{\bibinfo{person}{Zeynep Tufekci} {and}
  \bibinfo{person}{Christopher Wilson}.} \bibinfo{year}{2012}\natexlab{}.
\newblock \showarticletitle{Social media and the decision to participate in
  political protest: Observations from Tahrir Square}.
\newblock \bibinfo{journal}{\emph{Journal of communication}}
  \bibinfo{volume}{62}, \bibinfo{number}{2} (\bibinfo{year}{2012}),
  \bibinfo{pages}{363--379}.
\newblock


\bibitem[\protect\citeauthoryear{Von~Busch}{Von~Busch}{2022}]%
        {von2022making}
\bibfield{author}{\bibinfo{person}{Otto Von~Busch}.}
  \bibinfo{year}{2022}\natexlab{}.
\newblock \bibinfo{booktitle}{\emph{Making Trouble: Design and Material
  Activism}}.
\newblock \bibinfo{publisher}{Bloomsbury Publishing}.
\newblock


\bibitem[\protect\citeauthoryear{Wu, Lee, Chang, and Liang}{Wu
  et~al\mbox{.}}{2013}]%
        {wu2013current}
\bibfield{author}{\bibinfo{person}{Hsin-Kai Wu}, \bibinfo{person}{Silvia Wen-Yu
  Lee}, \bibinfo{person}{Hsin-Yi Chang}, {and} \bibinfo{person}{Jyh-Chong
  Liang}.} \bibinfo{year}{2013}\natexlab{}.
\newblock \showarticletitle{Current status, opportunities and challenges of
  augmented reality in education}.
\newblock \bibinfo{journal}{\emph{Computers \& education}}
  \bibinfo{volume}{62} (\bibinfo{year}{2013}), \bibinfo{pages}{41--49}.
\newblock


\bibitem[\protect\citeauthoryear{Wylie}{Wylie}{2020}]%
        {wylie2020promise}
\bibfield{author}{\bibinfo{person}{Alison Wylie}.}
  \bibinfo{year}{2020}\natexlab{}.
\newblock \showarticletitle{The promise and perils of an ethic of stewardship}.
\newblock In \bibinfo{booktitle}{\emph{Embedding ethics}}.
  \bibinfo{publisher}{Routledge}, \bibinfo{pages}{47--68}.
\newblock


\bibitem[\protect\citeauthoryear{Zhao, Bennett, Benko, Cutrell, Holz, Morris,
  and Sinclair}{Zhao et~al\mbox{.}}{2018}]%
        {zhao2018enabling}
\bibfield{author}{\bibinfo{person}{Yuhang Zhao}, \bibinfo{person}{Cynthia~L
  Bennett}, \bibinfo{person}{Hrvoje Benko}, \bibinfo{person}{Edward Cutrell},
  \bibinfo{person}{Christian Holz}, \bibinfo{person}{Meredith~Ringel Morris},
  {and} \bibinfo{person}{Mike Sinclair}.} \bibinfo{year}{2018}\natexlab{}.
\newblock \showarticletitle{Enabling people with visual impairments to navigate
  virtual reality with a haptic and auditory cane simulation}. In
  \bibinfo{booktitle}{\emph{Proceedings of the 2018 CHI conference on human
  factors in computing systems}}. \bibinfo{pages}{1--14}.
\newblock


\bibitem[\protect\citeauthoryear{Zuckerman}{Zuckerman}{2007}]%
        {zuckerman_2007}
\bibfield{author}{\bibinfo{person}{Ethan Zuckerman}.}
  \bibinfo{year}{2007}\natexlab{}.
\newblock \bibinfo{title}{The connection between cute cats and web censorship}.
\newblock
\newblock
\urldef\tempurl%
\url{http://ethanz.wpengine.com/2007/07/16/the-connection-between-cute-cats-and-web-censorship/}
\showURL{%
\tempurl}


\end{thebibliography}

\newpage

\onecolumn

\appendix

\section*{Appendix}

%\includepdf[pages=-]{Interview Questions [Supplemental Material].pdf}

\section{Interview Questions}

\begin{enumerate}

\subsection{Introduction and Background}

\item Can you provide a brief description of yourself and the work you do?
\item Tell me more about tools and techniques you used in your work.
\item Is your work intended to provide a social contribution? If so, why/how?
\begin{enumerate}
\item Would you say your work is connected to ‘activism’? Why or why not?
\end{enumerate}
\item On your (website, or other), I noticed you’ve described yourself as (an AR activist, or other). Could you explain what that means?

\subsection{Decisions around using AR for their project}

\item I’d like to learn more about <insert project name>. Can you tell me about it?

\begin{enumerate}
\item What did you do?
\item What was the goal of the project?
\item What audience were you trying to reach (if any)?
\item What is the history of this project?
\end{enumerate}

\item How did you incorporate AR in this project?

\begin{enumerate}
\item What tools did you use?
\item Why did you use these tools?
\item How was the process for learning to use these tools?
\end{enumerate}

\item Why did you decide to use AR for this project?  
\item Did you use any other technologies besides AR?

\begin{enumerate}
\item Why or why not?
\item Were you considering using any other technologies? If yes, why did you decide not to use them?
\end{enumerate}

\subsection{How AR influenced the project}

\item What would you change about your project, if anything?
\item What did you like about using AR, if anything?
\item What didn’t you like about using AR, if anything?
\begin{enumerate}
\item Did you experience any challenges in using AR for your project?
\end{enumerate}

\item Was there anything you wish you could have done with AR for this project?
\begin{enumerate}
\item Why?
\item What barriers did you experience to doing that?
\end{enumerate}

\item What impact do you believe this project had on <insert audience they described above>?
\begin{enumerate}
\item Were you able to achieve your goal of <insert goal described above>?
\item (If no) What do you think could have helped you achieve your goal?

\end{enumerate}

\item How did AR affect the project’s ability to have this impact, if at all?
\begin{enumerate}
\item (alternatively) Do you think the project could have had this impact without AR? 

\item Why or why not?
\end{enumerate}

\subsection{Broader experience with AR
}

\item Have you used AR in any other project?

\begin{enumerate}
\item Why or why not?
\item (if not) Have you considered using AR in another project? Why or why not?
\item Would you consider using it in future projects? Why or why not?
\end{enumerate}

\item Do you know any other activists who are using AR?

\begin{enumerate}
\item If yes, please explain.
\item What do you think about their work?
\item What do you think about how they used AR?

\end{enumerate}

\end{enumerate}

\end{document}